\documentclass[11pt]{article}

\usepackage[utf8]{inputenc}
\usepackage[T1]{fontenc}
\usepackage[english]{babel}
\usepackage[margin=2.5cm]{geometry}

\usepackage{amsmath}
\usepackage{amssymb}
\usepackage{physics}
\usepackage{graphicx}
\usepackage{xcolor}
\usepackage{siunitx}
\usepackage{ulem}
\usepackage{caption}
\usepackage{subfigure}
\usepackage{hyperref}
\hypersetup{colorlinks=true, linkcolor=blue, citecolor=blue, urlcolor=blue}

\usepackage{listings}
\lstdefinestyle{mystyle}{
    language=Python,
    basicstyle=\ttfamily\footnotesize,
    keywordstyle=\color{magenta},
    commentstyle=\color{gray},
    stringstyle=\color{red},
    numbers=left,
    numberstyle=\tiny\color{gray},
    stepnumber=1,
    breaklines=true,
    frame=single
}
\begin{document}

\title{Pressure and asymmetry govern the shape and stiffness of inflatables}

\author{Nathan Vani$^{1,2,\ast}$, Tom Joblin$^{1}$, Alejandro Ibarra$^{1,3,4}$, Jos\'e Bico$^{1}$, \'Etienne Reyssat$^{1}$, Beno\^it Roman$^{1}$}

\date{}

\maketitle

\vspace{-6mm}
\begin{center}
\begin{minipage}{0.92\textwidth}
\small
$^{1}$Physique et M\'ecanique des Milieux H\'et\'erog\`enes, CNRS, ESPCI Paris, Universit\'e PSL, Sorbonne Universit\'e, Universit\'e Paris Cit\'e, 7 quai Saint-Bernard, 75005 Paris, France\\
$^{2}$Institute of Physics, University of Amsterdam, Amsterdam, the Netherlands\\
$^{3}$Experimental Soft Matter Physics Group, Department of Physics and Materials Science, University of Luxembourg, Luxembourg L-1511, Luxembourg\\
$^{4}$Institute for Advanced Studies, University of Luxembourg, Belval Esch-sur-Alzette, Luxembourg\\
$^\ast$Corresponding author: nathan.vani@ens-paris-saclay.fr

\end{minipage}
\end{center}
\vspace{4mm}

\begin{abstract}

Inflatables made of thin sheets constitute a lightweight, scalable alternative to conventional soft robots. Since sheets are essentially inextensible while offering low resistance to bending, the shape of a straight tube should be trivially set by volume maximization. We show that networks of parallel tubes made from two sheets differing in stiffness defy this expectation as their global shape is governed by the binding angle at the junctions of adjacent tubes. Through this angle, the stiffness asymmetry induces a pressure-dependent curling and stiffening of the networks. Modeling a tube cross-section as two coupled rods, we quantitatively describe the geometry and mechanics of this new class of inflatables. Our model captures unexpected mechanical features such as a stiffness scaling as the square root of pressure and a contact-induced stiffening between neighboring tubes -- challenging common assumptions on thin-sheet inflatables. Unlike prior work restricted to the high-pressure regime, the pressure-dependent description further enables multiprogrammable control over a continuous range of curvatures. Discussing a variety of examples, we finally show that networks of asymmetric tubes are a versatile platform for functional shape-morphing objects.

\end{abstract}

\section{Introduction}

Soft robots are traditionally made of pneumatic elastomer networks which undergo significant stretching during inflation~\cite{gorissen2017elastic, whitesides2018soft}. The most common example is the gripper which uses fingers actuated in bending through differential stretching -- akin to Timoshenko's bilayer~\cite{timoshenko1925analysis} -- allowing precise, pressure-dependent control over geometry~\cite{shepherd2011multigait, jones2021bubble}. Elastomer-based actuators are nonetheless relatively bulky and cannot sustain large loads and high inflating pressures due to material limitations~\cite{ogden1972deformation}. Inflatables made of two bonded thin sheets offer an exciting alternative as they are lightweight, load-bearing, and easier to manufacture. First studied for aerospace applications~\cite{jenkins2001gossamer}, thin-sheet inflatables were reintroduced as soft actuators by Niiyama \textit{et al.}~\cite{niiyama2015pouch} and popularized by Ou \textit{et al.}'s \textit{aeromorphs}~\cite{ou2016aeromorph}, whose accessible fabrication method attracted a wide community spanning human-computer interaction~\cite{khin2017fabric, liang2018design, lu2019millimorph, arfaee2023modeling, yang2024snapinflatables}, computer graphics~\cite{panetta2021computational, ren2024computational, he2025matairials, mirkin2025programming}, and mechanical engineering~\cite{siefert2020geometry, zhang2023modular, andrade2023fabric, broshkevitch2025programmable, yao2026hybrid}.

Mechanically, thin-sheet inflatables differ widely from elastomer-based ones as the sheets are stiff enough not to stretch significantly yet thin enough to bend easily~\cite{siefert2020programming}. Considering a tube made of sheets of thickness $t$, Young's modulus $E$, and width $W$, this amounts to the conditions on the inflating pressure: $Et^3/W^3 \ll P \ll Et/W$. Working in this doubly asymptotic inextensible regime allows for straightforward shape programming through volume maximization, either using kinks as hinges~\cite{ou2016aeromorph, yang2024snapinflatables, yao2026hybrid, yang2026reconfigurable}, overcurvature in bent tubes~\cite{siefert2019programming, andrade2023fabric}, buckling due to anisotropic shrinking~\cite{panetta2021computational}, or differential contraction in layered tubes~\cite{mirkin2025programming}. Though few studies leave this regime~\cite{he2023cross, chandler2025mechanics}, this approach is severely limited as programming becomes binary: balloons are considered either deflated or fully inflated, with pressure playing no role in the final geometry. The associated designs are especially limiting for soft robotics applications which require pressure-dependent actuation, e.g. to build an artificial heart~\cite{vis2022ongoing, rogatinsky2023multifunctional, arfaee2025soft}.

\begin{figure}
  \includegraphics[width=1\linewidth]{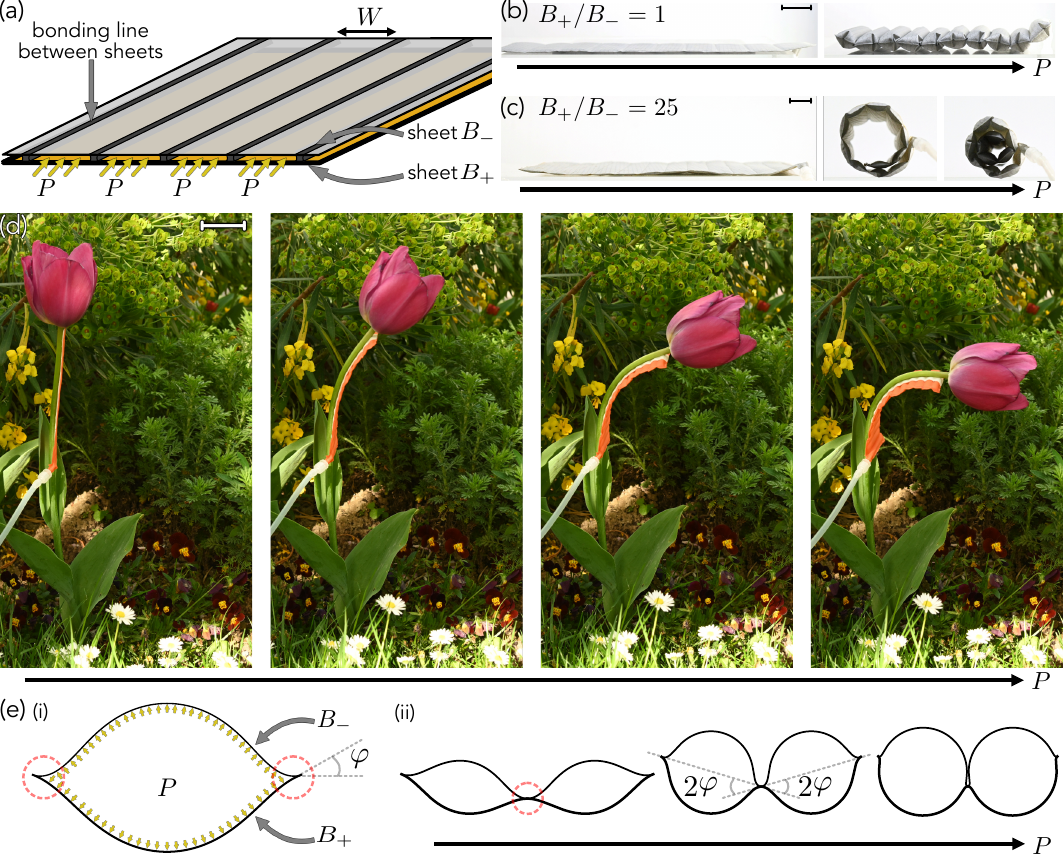}
  \caption{Networks of asymmetric tubes. 
  (a) Networks of parallel tubes of width $W$ are manufactured by soldering two sheets of bending stiffness $B_+$ and $B_-$, and inflated at pressure $P$. 
  (b-c) Inflation of a symmetric (b) network which remains overall flat and an asymmetric (c) network which curls towards the less stiff sheet. Scale bar is \SI{2}{\centi\meter}. (d) A symmetric network is taped to the stem of a flower and inflated at an increasing pressure. The stem creates a stiffness asymmetry causing the system to curl elastically. Scale bar is \SI{4}{\centi\meter}.
  (e) (i) The cross-section of a tube modeled as two bound planar rods of length $W$ and stiffness $B_+$ and $B_-$, loaded by a uniform pressure $P$. Red circles highlight the junctions in which boundary layers develop at larger pressures. (ii) As the pressure increases, two adjacent tubes deviate by an increasingly large angle 2$\varphi$, up to the contact between the tubes.}
  \label{fig:introduction}
\end{figure}

In this study, we consider networks of tubes made from two sheets of bending stiffness $B_+$ and $B_-$ per unit length ($B\sim Et^3$), bonded in a planar manner to form a network of straight parallel tubes of width $W$, as shown in Fig.~\ref{fig:introduction}a. 
For two identical sheets, the network remains macroscopically flat upon inflation, with mostly circular cross-sections (Fig.~\ref{fig:introduction}b) -- an observation well understood through volume maximization. However, if both sheets differ in stiffness (Fig.~\ref{fig:introduction}c), the network curls toward the less stiff sheet with an amplitude that strongly depends on the actuating pressure $P$. The resulting curvature is particularly large, only limited in practice by tubes coming into contact. The phenomenon is illustrated by taping a symmetric network of tubes to the stem of a flower before pressurizing the system (Fig.~\ref{fig:introduction}d). The stem induces a stiffness asymmetry causing the flower to bend continuously as the applied pressure is increased. The phenomenon can neither be understood nor modeled by considering the doubly asymptotic regime explored in previous studies.

A model of the cross-section of a tube is shown in Fig.~\ref{fig:introduction}e (i). The two asymmetric sheets, here considered asymmetric, are modeled as two inextensible bending rods, tangent at their extremities so as to introduce a cusp in curvature. In the case of $B_+/B_- > 1$, the coupling results in a non-zero junction angle $\varphi$ between the central axis of the tube and the direction of the shared end. Such finite junction angles can be found on both sides of each soldering line, and their accumulation results in the overall bending of the whole structure -- as seen in Fig.~\ref{fig:introduction}e (ii).

In the following, we first discuss the geometry of a single tube, presenting the cross-section model. We then extend the analysis to account for contact between adjacent tubes and discuss the stiffness of inflatable tubes. The mechanics of asymmetric tubes highlights how small structural elements, here bending boundary layers, can govern the large-scale response of thin-sheet inflatables. We finally show that asymmetric thin-sheet networks offer a promising platform for shape-morphing and pave the way toward soft robotics-ready pneumatic actuators.

\newpage
\section{Results}

\subsection{Shape of a cross-section}

\begin{figure}
  \includegraphics[width=\linewidth]{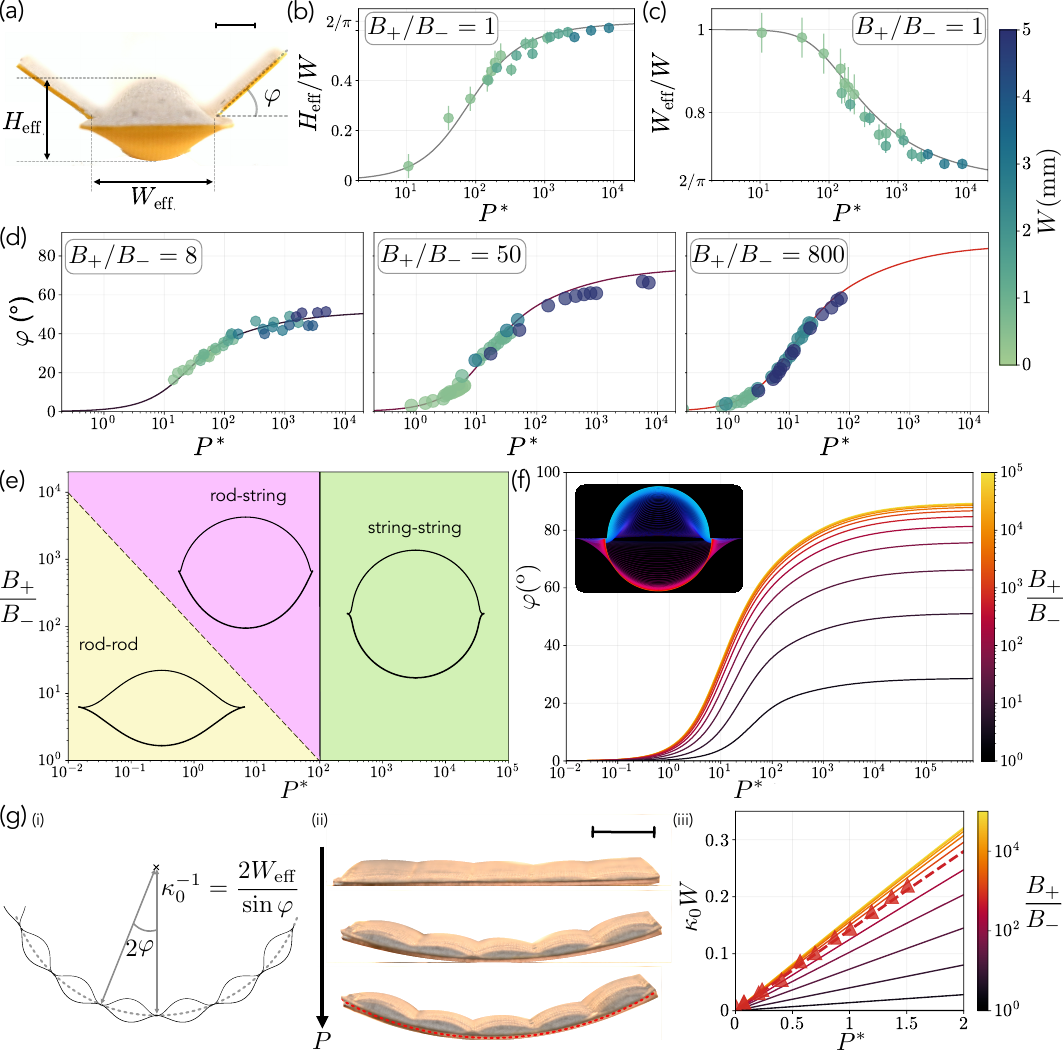}
  \caption{Geometry of a tube. (a) An asymmetric tube ($B_+/B_- = 10$) is inflated to a pressure $P$, resulting in an effective contraction of its width $W_{\mathrm{eff}}$ and increase of its height $H_{\mathrm{eff}}$. Small fins are attached to its sides to visualize the junction angle $\varphi$. (b-d) Evolution of the geometrical parameters $H_{\mathrm{eff}}$, $W_{\mathrm{eff}}$ and $\varphi$ as a function of $P^*$ according to experiments (markers), compared to predictions from the cross-section model (lines), for various stiffness asymmetries (legend) and tube widths (colorbar). (e) Configuration diagram of a cross-section as a function of the dimensionless pressure $P^*$ and stiffness asymmetry $B_+/B_-$, distinguishing whether each rod acts as a string (bending stiffness negligible in the bulk) or a rod (bending stiffness significant in the bulk). (f) Evolution of the junction angle $\varphi$ as a function of $P^*$ and $B_+/B_-$ (colorbar) as predicted by the cross-section model: the more asymmetric, the larger the rotation for a given dimensionless pressure. Inset: visualization from the model of a cross-section as it inflates. (g) The curvature of a network of tubes $\kappa_0$ is derived from the model of a single cross-section (i) with corresponding experiments shown in (ii). Both experimental and numerical data are plotted in (iii). Scale bar is \SI{4}{\centi\meter}. Markers correspond to low pressure experiments at $B_+/B_- = 800$, in good agreement with the corresponding prediction plotted as a dashed line.} 
  \label{fig:cross-section}
\end{figure}

We consider here a single slender tube, aiming to characterize its geometry along the bulk of its length. Such a tube is shown in Fig.~\ref{fig:cross-section}a with fins attached along its sides to visualize the junction angle $\varphi$. We also define  its effective width $W_{\mathrm{eff}}$ and height $H_{\mathrm{eff}}$. These three observables
depend on the four parameters of the system: ($P$, $W$, $B_+$, $B_-$) that can be described as a function of two dimensionless parameters. A natural parameter is the stiffness ratio $B_+/B_-$. The second can be understood as a dimensionless pressure:

\begin{equation}
    P^* = \frac{PW^3}{B_+},
\end{equation}

\noindent which correspond in scaling to the ratio of the typical torque applied by pressure over a sheet ($\sim PW^2$) to the torque required to induce a curvature of typical size $1/W$ in the stiffer sheet ($\sim B_+/W$). Note that we have chosen $B_+$ over $B_-$ here as the stiffer sheet sets the range of working pressures in practice.

For symmetric tubes, we report the evolution of $W_{\mathrm{eff}}$ and $H_{\mathrm{eff}}$ rescaled by the initial tube width $W$ in Fig.~\ref{fig:cross-section}b-c as a function of $P^*$ for symmetric tubes of various widths (colorbar). The collapse of the data on a single curve confirms that $P^*$ is the single dimensionless parameter in the symmetric configuration $B_+/B_-=1$. The usual binary behavior is recovered through the data: when deflated, the cross-sections is flat ($W_{\mathrm{eff}} = W$, $H_{\mathrm{eff}} \simeq 0$), and when highly inflated the section shape trends toward a circle ($W_{\mathrm{eff}} = H_{\mathrm{eff}} = 2W/\pi$). We note that the inflating pressure required to reach this regime is quite large ($P^* \gg 10^3$), a condition rarely met in practice as soft robots typically involve inflatable elements with $W \sim $ \SI{1}{\centi\meter} and $B \sim10^{-3}\,$mN.m., thus requiring $P \gg$ \SI{10}{\bar}. The behavior is therefore set by the intermediate pressure regime, where the cross-sectional geometry varies continuously with $P^*$ and cannot be reduced to its asymptotic limit.

Regardless of the pressure, we observe a non-zero junction angle $\varphi$ whenever $B_+/B_- > 1$. We plot the evolution of $\varphi$ as a function of $P^*$ in Fig.~\ref{fig:cross-section}d for three ratios of asymmetry. The junction angle strongly depends on $B_+/B_-$ -- at a given dimensionless pressure, the larger the asymmetry, the larger the angle. A finite, non-zero value of $\varphi$ is found even at large inflating pressures when the cross-section approximates a circle.

We propose to analyze the system through a single cross-section, considering the tube invariant by translation along its length. Fig.~\ref{fig:cross-section}e shows a configuration diagram of the cross-section as a function of $P^*$ and $B_+/B_-$. For large $P^*$, pressure dominates bending in both rods, each one taking the shape of a half-circle -- this regime is named ‘string-string' as the bulk shape does not see any influence of their bending stiffness. The relevant dimensionless pressure for the softer rod is $PW^3/B_-$, which corresponds to the product of the two dimensionless parameters considered in the diagram. Two additional regimes are observed for lower values of $P^*$: ‘rod-string', where only the softer rod takes the shape of an arc of circle, and ‘rod-rod', where both bending stiffnesses are significant. The cross-sections shown are obtained through numerical resolution of the system. Each sheet is modeled as a rod which deforms in bending only, with assumptions of inextensibility and linear elasticity. The model leads to a system of 12 differential equations, 6 per rod. The rods are coupled at their ends so as to point in the same direction with compatibility of positions, forces and torques. The equations and their numerical resolution are detailed in Supplementary Section C. Supplementary video MovieS3 and the inset of Fig.~\ref{fig:cross-section}f show the evolution of the cross-section with increasing pressure.

From the simulations, we extract the geometrical variables, as plotted for $\varphi$ in Fig.~\ref{fig:cross-section}f. The model predictions agree quantitatively with experiments, as shown in Fig.~\ref{fig:cross-section}b-d. Further experimental data are discussed in Supplementary section C -- validating the model in width, height, and junction angle for other stiffness ratios.

The model further confirms the relevance of $\varphi$ as a marker of bending stiffness. This angle stems from the coupling condition of the sheets, a direct consequence of the fabrication method of thin-sheet inflatables. Whereas previous studies~\cite{ou2016aeromorph, siefert2020programming, gao2020shape, panetta2021computational, gao2023pneumatic, ren2024computational, he2025matairials} considered the bending stiffness to be irrelevant to the geometry of inflatables, we observe here that the binding angle makes the ratio of stiffness relevant to the final shape regardless of the applied pressure.

The phenomenon is best illustrated in the ‘string-string' limit in which bending boundary layers develop on each side of the junction. In this regime, tensile forces in the sheets, scaling as $PW$, dominate over the applied pressure around the relatively small junction whose typical length scales as $\sqrt{B_\pm/PW}$. In the vicinity of the junction, one can thus model the connection between the sheets as two rods subject only to tension at their ends. This type of boundary layer occurs in numerous systems of coupled slender structures~\cite{vani2025asymmetric}. Modeling the junction as two coupled \textit{elastic\ae} gives the asymptotic value of the angle for large pressures:

\begin{equation}
    \sin \varphi \left(P^* \to +\infty\right) = \frac{B_+/B_- - 1}{B_+/B_- + 1},
%    \sin \varphi \left(P^* \to \infty\right) = \frac{B_+ - B_-}{B_+ + B_-},
\end{equation}

which corresponds to the plateau value observed numerically. It is important here to highlight that the curling of asymmetric networks of tubes is a local effect due to torque balance at the junction of sheets -- an area which coincides with the junction of consecutive tubes. The effect differs from the curling induced by differential stretching as used classically in shape-morphing systems~\cite{klein2007shaping, reyssat2009hygromorphs, pezzulla2016geometry}. Asymmetric junctions allow for a pressure-depending tuning of the macroscopic curvature of networks of tubes thanks to discrete local junctions. Bending stiffness thus matters regardless of the pressure regime contrary to previously held assumptions.

The modeling of a single cross-section further predicts the curvature $\kappa_0$ of a network of asymmetric tubes from the effective width $W_\mathrm{eff}$ and the junction angle $\varphi$. As sketched in Fig.~\ref{fig:cross-section}g (i), the curvature is:

\begin{equation}
    \kappa_0 = \frac{\sin \varphi}{2W_{\mathrm{eff}}}.
\end{equation}

We show the inflation of such a network in this regime in Fig.~\ref{fig:cross-section}g (ii), and plot experimental data along numerical predictions in (iii). We again find a good agreement with our model. In addition to the numerical predictions, we provide approximate expressions for the junction angle at low pressures in Supplementary Section C in both the ‘rod-rod' and ‘rod-string' regimes. Interestingly, straightforward linearization fails to capture the evolution of $\varphi$ in this system due to the coupled effect of inextensibility and the compatibility conditions at the junction -- a theoretical discussion beyond the scope of the present paper.

Note that the present discussion holds within the assumptions of negligible stretching of the sheets, slenderness ($P \ll B_\pm / W t_\pm^2$), and linear elasticity, the latter breaking down when the stress at the junction exceeds either the yield strength or the linear elastic limit of the sheets~\cite{vani2025asymmetric}. Beyond these thresholds, the junction angle can follow a variety of behaviors depending on the sheet properties, as tension rather than bending governs the junction area. We discuss such regimes in Supplementary section D. In practice, for typical thin-sheet inflatables, contact between neighboring tubes occurs before any of these thresholds is reached, as discussed in the following section. The resulting counter-torque at the junction decreases $\varphi$, such that single cross-section model over-predicts the curvature of arrays at large pressures.

\newpage
\subsection{Contact between neighboring tubes}

\begin{figure}
  \includegraphics[width=\linewidth]{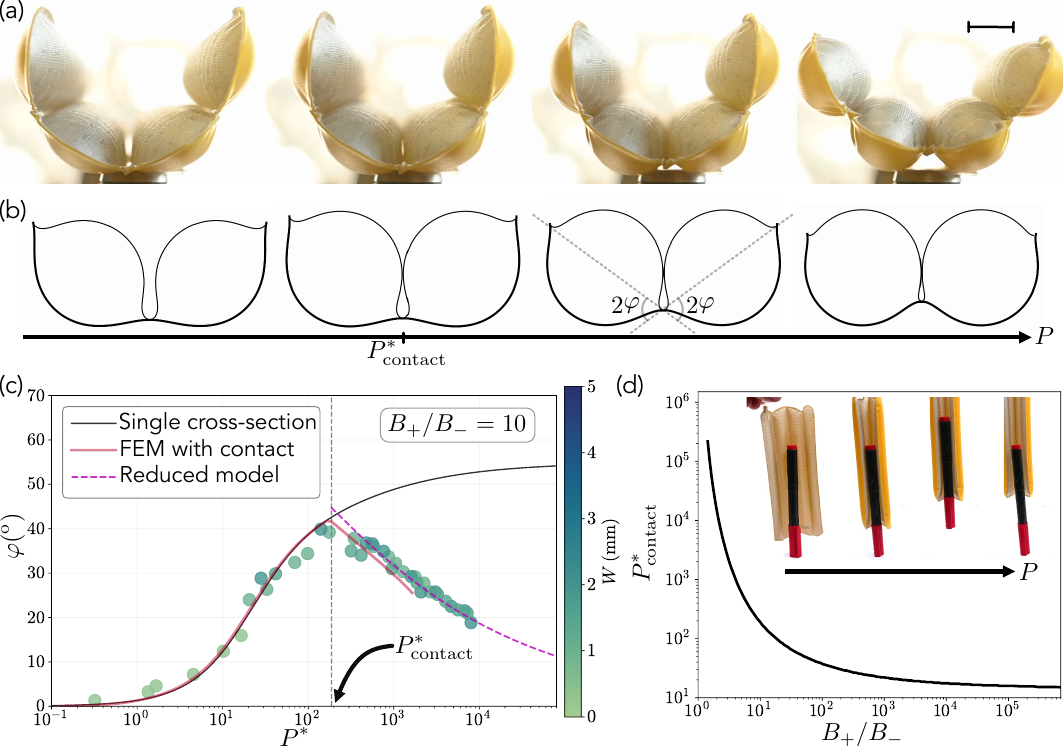}
  \caption{Interaction between consecutive tubes. (a-b) Network of tubes as the pressure is increased, experiments (a) and FEM simulations (b). Scale bar is \SI{1}{\centi\meter}. At pressure $P^*_{\mathrm{contact}}$, contact occurs between the less stiff sheets of neighboring tubes and the trend reverses as the structure uncurls. %with decreasing junction angle $\varphi$.
  (c) Evolution of angle $\varphi$ measured in neighboring tubes (markers), alongside predictions from the single cross-section model (black line), FEM simulations accounting for contact (red line), and prediction from the reduced model Eq.~\eqref{eq:reduced_model} (dashed purple line). The single cross-section model only predicts the onset of contact but fails to capture the trend starting at $P^*_{\mathrm{contact}}$. (d) Evolution of $P^*_{\mathrm{contact}}$ as a function of the stiffness asymmetry $B_+/B_-$. Inset: the non-monotonic bending of a network is used to grab and then release a pen with a monotonic increase of the actuating pressure.}
  \label{fig:contact}
\end{figure}

The cross-section model previously detailed is blind to interactions between neighboring tubes. Since the junction angles are particularly large at high pressures, contact between neighbour tubes with nearly circular cross-sections occur away from the junction area. As the previous discussion resulted from experiments on single tubes, we now turn to the inflation of networks of tubes as shown in Fig.~\ref{fig:contact}a. At a dimensionless threshold pressure $P^*_{\mathrm{contact}}$, the less stiff sheets of neighboring tubes touch, and the trend in curvature reverses as the network starts to uncurl. The tubes appear to push against each other in the post-contact regime, applying a counter-torque at the junction. To account for contact, we model the system using finite-element methods (FEM) software COMSOL Multiphysics. Cross-sections of two tubes modeled with frictionless contact are shown in Fig.~\ref{fig:contact}b, displaying a similar behavior. The FEM model is based on shell theory and therefore captures both bending and in-plane stretching of the plate. The structure is represented by two independent surfaces connected along three lines to account for the seams: two along the edges and one central seam, which is additionally constrained. Contact is modeled as frictionless. To improve computational efficiency, the contact detection algorithm is restricted to surfaces with lower bending stiffness, where contact is expected to occur \textit{a priori}.
 
We report experimental data in Fig.~\ref{fig:contact}c, along the model for a single cross-section (black line). This model can be used to predict the threshold pressure $P^*_{\mathrm{contact}}$ using the symmetric properties of the two tubes system (dashed black line), as well as the angle at contact $\varphi_{\mathrm{contact}} = \varphi\left(P^*_{\mathrm{contact}}\right)$. We discuss these geometrical considerations in Supplementary section E. The model nonetheless fails to capture the evolution of the junction angle after contact, while FEM simulations predict correctly the behavior (red line) with a sublinear decrease of the junction angle with the pressure. As the collapse of the data still holds, bending still appears as the primary deformation mechanism here.

The post-contact evolution of $\varphi$ is also captured by a simple geometrical model which considers the cross-sections as nearly circular and contact as punctual, so that $\varphi$ is set only by the geometry of two tangent circles. The contact is thus supposed to apply a torque on the junction with a typical lever arm of the size of the boundary layer. Full description is provided in section E of the Supplementary. The model leads to the description in the post-contact regime:

\begin{equation}
    \cos \varphi = \frac{1}{1 + \sqrt{\pi^3/P^*}},
\label{eq:reduced_model}
\end{equation}

which agrees well with experiments and numerical predictions without any fitting parameter (purple dashed line in Fig.~\ref{fig:contact}c). In this reduced model, the angle vanishes in the limit of infinite pressure -- a prediction that could not be verified experimentally as the soldering lines failed before large enough pressures could be reached. The limit appears reasonable: for purely circular cross-section, contact would occur only at the junction point leading to $\varphi=0$.

Ultimately, contact between consecutive tubes acts as a maximal limitation for the curling of arrays of asymmetric tubes. In Fig.~\ref{fig:contact}d, we plot the pressure of first contact based on the stiffness asymmetry as predicted by the single cross-section model. Contact occurs in practice in the range of $P^*$ between 20 and 200, which corresponds to a maximal achievable angle of around 45°. The associated decrease of curvature in the junction area of the weaker rod delays the breaking of slender linear elasticity assumptions. Considering the large dimensionless pressures, this is surprising from boundary layer analysis only.

The inset of Fig.~\ref{fig:contact}d demonstrates how the post-contact regime can be used for shape-programming. The non-monotonic relation between actuating pressure and network curvature is used in a network of tubes to grab and then release an object while the controlling pressure is increased monotonically. Such nonlinearities could thus be used to engineer complex responses from simple actuation.

\newpage
\subsection{Bending rigidity of tube networks}

We now turn to the bending rigidity of tube networks along the direction of curling, i.e. orthogonally to the bonding lines. Though numerous studies have been dedicated to the stiffness of tubes along their lengths~\cite{comer1963deflections, webber1982deflections, main1994load, wielgosz2002deflections, thomas2013exact, thomas2018mechanics} -- see~\cite{nguyen2015inflation} for a short review, little attention has been paid to the transverse direction. In models used to design thin-sheet inflatables with Gaussian curvature, the transverse stiffness is generally considered to be governed by the compliance of soldering lines, and thus independent of the inflating pressure and small enough to be insignificant~\cite{panetta2021computational, ren2024computational, he2025matairials, gao2020shape}.

\begin{figure}
  \includegraphics[width=\linewidth]{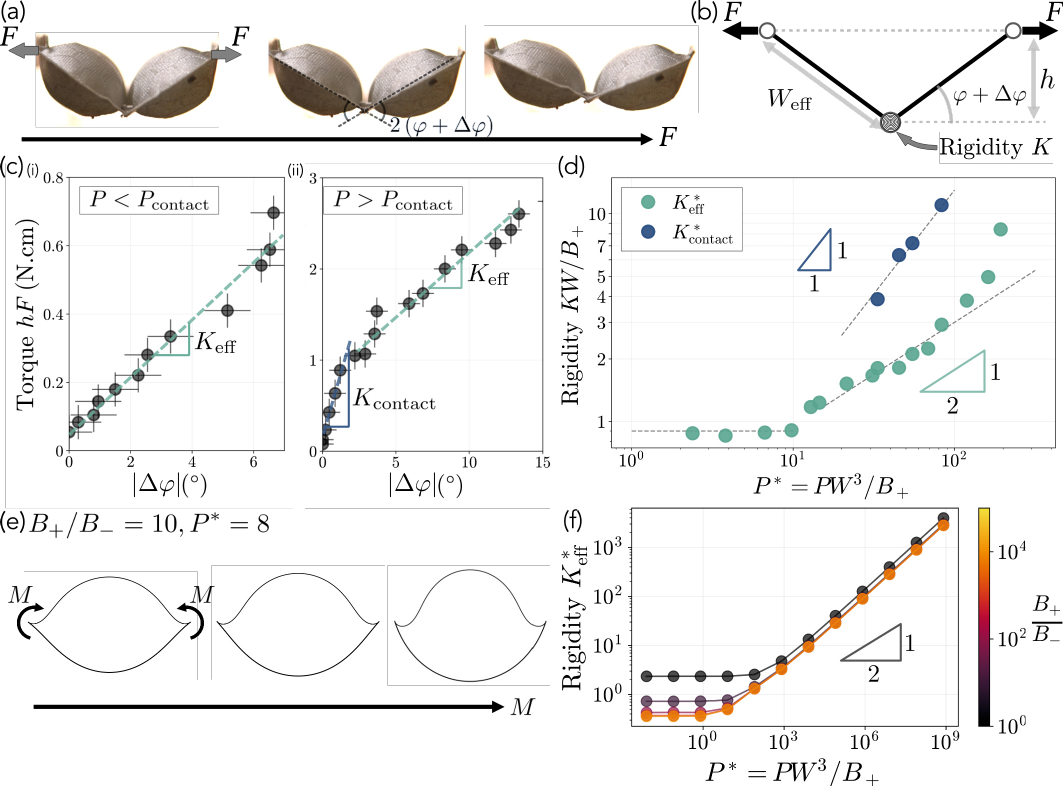}
  \caption{Rigidity of networks of tubes. (a) Characterization of the rotational stiffness: two neighboring tubes are loaded with an increasing pulling force. Here and in the next plots $B_+/B_-=50$ and $W=$\SI{2}{\centi\meter}. (b) As the deformation is localized at the junction of the tubes, we model the system as two rigid bars of length $W_\mathrm{eff}$ connected by a pivot of stiffness $K$ loaded by a torque $hF$, leading to a variation $\Delta \varphi$ of the junction angle. (c) Experimental results: variation of $hF$ with the absolute value of $\Delta \varphi$. At small loads, the response is linear if the tubes are not in contact (i), or piecewise linear if they initially are (ii) as the contact stiffens the system until contact is broken. We measure the corresponding coefficients $K_\mathrm{eff}$ and $K_\mathrm{contact}$. (d) Evolution of the rigidity coefficients made dimensionless by $B_+/W$ as a function of $P^*$. At low pressures, the contactless stiffness $K_\mathrm{eff}$ remains constant and the stiffness at contact $K_\mathrm{contact}$ is not defined for $P^* < P^*_{contact}$. $K_\mathrm{eff}$ then varies as $\sqrt{P^*}$ and $K_\mathrm{contact}$ as $P^*$. (e) We model the rotational stiffness by applying point torques to the junctions in the single cross-section model. (f) Evolution of the contactless rigidity as a function of $P^*$ and $B_+/B_-$ (colorbar) as predicted by the cross-section model.}
  \label{fig:stiffness}
\end{figure}

% We aim to characterize the first-order mechanical response of series of tubes.
In Supplementary Section F, we show experimentally that the stiffness at low inflating pressures ($P^* < 10$) is governed by the stiffness of the sheets with no discernible influence of the pressure. More interestingly, we probe the regime of larger dimensionless pressures with a second experiment shown in Fig.~\ref{fig:stiffness}a: two tubes are pulled apart in opposite directions at their free ends. In the regime considered, neither the cross-section nor the sheets stretch significantly under the newly applied load -- the deformation is localized at the junction between the tubes. The system is modeled as two rigid bars of length $W_\mathrm{eff}$, initially at an angle $\pi - 2\varphi$ for $F=0$. A hinge with a rotational stiffness $K$ corresponding to the boundary layer connects the bars and is subjected to a torque $M= Fh = FW_\mathrm{eff}\sin\left(\varphi  + \Delta \varphi\right)$ with $\Delta \varphi$ the variation in junction angle. We thereafter characterize the linear coefficient of the response $K=\partial M/ \partial \Delta\varphi$ between the applied torque and the angle variation.

The response depends on whether neighboring tubes are in contact at $F=0$. We plot the evolution of the applied torque with the variation of the junction angle in Fig.~\ref{fig:stiffness}c in both cases. For $P^* < P^*_{\mathrm{contact}}$, the observed response is linear, whereas it is piecewise linear for $P^* > P^*_{\mathrm{contact}}$. In the latter case, the two upper sheets are initially in contact which stiffens the system. The sheets separate after a threshold in torque, leading to a second linear regime corresponding to a lower rotation stiffness -- comparable to the pre-contact regime. We note $K_{\mathrm{eff}}$ the stiffness outside of contact -- the only coefficient for $P^* < P^*_{\mathrm{contact}}$ and the second one for $P^* > P^*_{\mathrm{contact}}$, and $K_{\mathrm{contact}}$ the stiffness in case of contact.

We report the evolution of both stiffnesses as a function of $P^*$ in Fig.~\ref{fig:stiffness}d. At low pressures, $K_\mathrm{eff}$ is independent of $P^*$, consistent with the bending stiffness of the sheet dominating the response. Strikingly, above $P^* \sim 10$, $K_\mathrm{eff}$ grows as $\sqrt{P^*}$. In this regime, most of the cross-section rotates as a rigid body and the deformation is localized to the coupled bending boundary layers at the junction of the sheets. The tension in the boundary layers scales as  $PW$, with their typical size $\ell$ scaling as $\sqrt{B_\pm/PW}$ which sets the typical hinge length. Stiffness thus grows as $K\sim B/\ell \sim \sqrt{B PW}$. A more complete description considering two coupled elastic\ae is derived in~\cite{vani2025stiffness}, showing that the expected scaling is $\sqrt{PW\left(B_+ + B_-\right)}$. The scaling is further confirmed by the model of the cross-section  modified to account for a point torque at the edge of the tube as shown in Fig.~\ref{fig:stiffness}e. The resulting rigidity for various stiffness ratios for increasing dimensionless pressures is shown in Fig~\ref{fig:stiffness}f, capturing the experimental trend.

We further report $K_\mathrm{contact}$ for pressures above $P^*_\mathrm{contact}$ (blue markers in Fig.~\ref{fig:stiffness}d). Contact stiffens the junction by a factor of about 2 relative to $K_\mathrm{eff}$ at comparable pressures, and the pressure dependence is steeper than $\sqrt{P^*}$. The lever arm, of the order of $\ell$, depends weakly on the inflating pressure whereas the force exerted on each other by the tubes scales as $P$, consistent with a near-linear scaling of $K_\mathrm{contact}$ with $P^*$. FEM simulations reproduce both stiffness trends quantitatively (Supplementary section F). A complete theoretical description of the contact-induced stiffening is beyond the scope of this paper -- however, this nonlinear stiffening appears promising for programmable soft actuators.

We have therefore identified two distinct pressure-dependent regimes for $P^* > 10$, with the stiffness either being governed by bending boundary layers with a scaling in $\sqrt{P^*}$, or being governed by contact between sheets with a linear scaling in $P^*$. These results directly challenge the usual assumption of pressure-independent transverse rigidity in the design of thin-sheet inflatables at the typical scale of soft robots. At very large dimensionless pressures, the rigidity induced by the boundary layers strongly increases and we still expect the bonding lines to be the most compliant elements of the system but this regime remains to be characterized. Further studies on the mechanics of networks of tubes are thus warranted, e.g. by accounting for the influence of extensibility outside of the first-order response~\cite{chandler2025mechanics} or the mechanical strength of the bonding lines~\cite{broshkevitch2025programmable} -- both of which carry a non-trivial pressure dependence that current models do not capture.

\newpage
\section{Applications}

We now present several applications of asymmetric tubular networks. Arrays with periodically alternating curvature are first modeled and demonstrated as lifters. We further explore the design space, showing that networks can be assembled with cuts to make kirigami-like tunable particle filters, and that arrangements of tubes gives control over the curvature along the inflatable. Such mechanisms are then used to design complex shapes with strategies based on origami and metric frustration. We conclude with a dynamic example of the activation of a snapping event.

\begin{figure}
  \includegraphics[width=\linewidth]{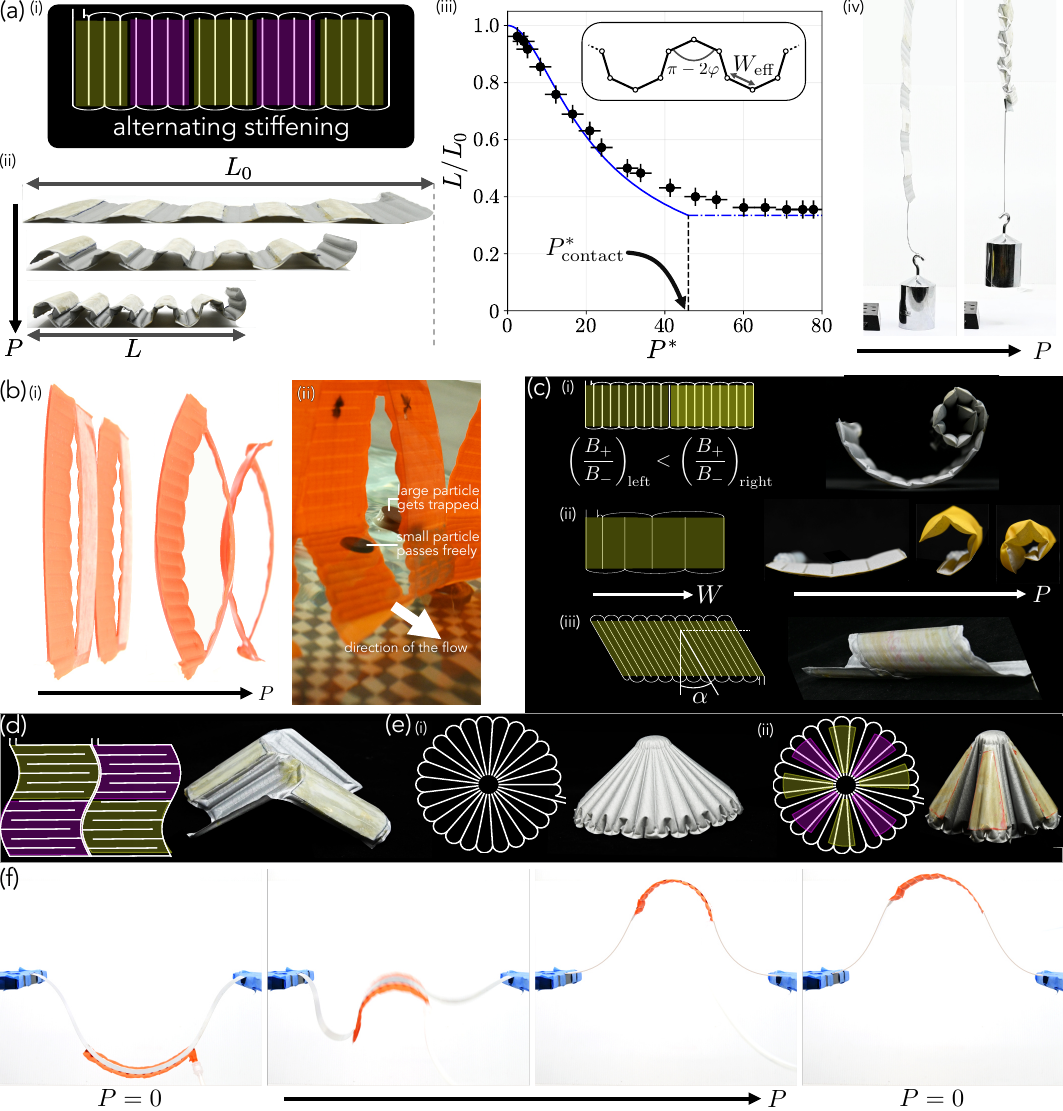}
  \caption{Uses of asymmetric networks of tubes -- tube widths are of the order of one centimeter, and inflating pressure of one bar. (a) (i) Pattern of a network with alternating recto (yellow) and verso (purple) stiffening patches applied. (ii) The resulting balloon
of length $L_0$ at $P=0$ shrinks macroscopically upon inflation due to out-of-plane undulations of its surface. (iii) Evolution of the shrinking ratio $L/L_0$ with $P^*$. The blue lines correspond to a macroscopic model as sketched in the inset, with $W_\mathrm{eff}$ and $\varphi$ as inputs from the cross-section model. (iv) The \SI{20}{\gram} inflatable is used to lift a \SI{1}{\kilo\gram} weight.  (b) Kirigami-like inflatable (i), used as a tunable filter for particles floating in a water channel (ii). (c) The curvature along an inflated network can be tuned by changing the stiffness asymmetry (i) or the tube width (ii). The object can also be programmed to curl along a helix (iii) by introducing an angle $\alpha$ between the direction of the tubes and the edge of the inflatable. (d) Unit cell of a curve fold Miura Ori origami, folded upon pressurization using alternating curved networks of tubes. (e) A geometrically frustrated inflatable shapes into a cone upon inflation with symmetric network (left). The shape of the cone is given by the shrinking ratio of tubes $2/\pi$~\cite{siefert2020programming}. Using alternating stiffening (right), the macroscopic shrinking ratio can be controlled to obtain a smaller apex angle. (f) A network of tubes is taped to a buckled beam. Pressurization leads to a snapping event.}
  \label{fig:application}
\end{figure}

Fig.~\ref{fig:application}a shows the design of a structure with varying curvature, here with a set value of stiffness asymmetry while alternating direction with a periodicity of three soldering lines. Upon inflation, the systems exhibits curvature with alternating sign, and the undulation along the surface leads to an important shrinking of the object along its length. We model this shrinking at the scale of the network using geometrical variables predicted by the cross-section model as inputs. Similarly to the model used for the transverse stiffness, tubes are considered as rigid bars of length $W_{\mathrm{eff}}$ connected by hinges of angle $\pi - 2\varphi$. The number of tubes in a given period is $n$. Adding the projections of each bar onto the symmetric axis gives the shrinking ratio:

\begin{equation}
    \frac{L}{L_0} = \frac{W_{\mathrm{eff}}}{nW}\cos\varphi\frac{\sin\left(n\varphi\right)}{\sin\varphi}.
\end{equation}

Fig.~\ref{fig:application}a (iii) shows a good agreement between experiments (markers) and this macroscopic model (blue line) for $n=3$. At the contact pressure, the structure stops shrinking -- the reduction in junction angle is compensated by the continued shrinking of individual tubes, leading to a plateau at the observed pressures. Such inflatables can be used as actuators to lift objects -- we show an example in Fig.~\ref{fig:application}a (iv) of a 20g inflatable lifting 50 times its own weight. A natural extension not explored here would be to determine the tensile stiffness of the structure by using the rotational stiffness of the tube junctions into the macroscopic model.

In Fig.~\ref{fig:application}b, we show a more complex structure combining alternating curvature directions and cuts to reproduce the behavior of a kirigami unit cell~\cite{isobe2016initial}. Upon inflation, the cuts open into apertures whose size is set by the curvature of the branches, and thus by the inflating pressure. Placed against the direction of a flow, the structure acts as a particle filter with tunable pore size for dilute to semi-dilute suspensions~\cite{guariguata2012jamming, vani2022influence}.

% Alternating the side on which stiffening patches are applied controls the direction of the curvature.
Considering the scaling in $P^*$ of geometrical parameters, it is possible to vary the amplitude of the curvature in a network by either varying the ratio of asymmetry along the structure, as in Fig.~\ref{fig:application}c (i), or by varying the widths of tubes, as in Fig.~\ref{fig:application}c (ii). By considering non-rectangular domains still filled with parallel tubes, we can obtain portions of cylinders that do not close onto themselves, such as a helix as in Fig.~\ref{fig:application}c (iii). Considering an angle $\alpha$ as sketched onto the soldering pattern, and a curvature $\kappa_0$, the radius of the helix is $\kappa_0^{-1}$, its pitch is $2\pi \kappa_0^{-1} \tan \alpha$, and its contraction rate along the longer edge of the pattern is $W_\mathrm{eff}/W \cos \alpha$. The control over the curvature can furthermore be used to obtain more complex surfaces. In Fig.~\ref{fig:application}d, we show curved fold based on Miura-Ori pattern actuated by inflation. Note that here the folding motion comes from the control of the surface itself rather than from the folding line~\cite{desimone2025mechanics}. Fig.~\ref{fig:application}e (i) shows an inflatable made from two identical sheets which inflates into a cone due to geometrical frustration~\cite{siefert2020programming}. The angle of the cone is $2 \arcsin \lambda$ with $\lambda$ the contraction rate, which corresponds to $W_\mathrm{eff}/W$ in the symmetric case, reaching a maximum of $2/\pi$ corresponding to an angle of \SI{80}{\degree}. Using the undulation structure previously discussed, we are able to obtain a cone with a sharper angle due to an improved effective shrinking in Fig.~\ref{fig:application}e (ii). The addition of stiffness asymmetry thus allows to reach larger Gaussian curvature normally not accessible using symmetric tubes in geometrically frustrated structures~\cite{ren2024computational}.

In Fig.~\ref{fig:application}f, a buckled sheet is affixed with a network of tubes -- the activation of which leads to a snapping event between buckled states~\cite{pandey2014dynamics}. The example hints at further developments that would account for the dynamics of pressurization in inflatables. Application could span from jumpers~\cite{gorissen2020inflatable} to frugal actuators~\cite{marzin2025augmented}.

Networks of tubes offer numerous further applications. We have shown in Fig.\ref{fig:introduction}d that an inert slender structure can be actuated through the simple adhesion of a symmetric tube network as a patch -- the asymmetry here arising from the addition of the stiffness of the object to the inflatable membrane. We show in Supplementary Section H (Fig. S15) a similar application with a tree branch -- suggesting uses as a scarecrow, or for snow removal and fruit harvesting. Controlling the motion of slender structures extends to human-made objects, with applications to morphable wings in small aircraft, tunable beams for civil engineering, or pressure-actuated cabinet doors. Furthermore, we demonstrate in Supplementary Section H that asymmetric tube networks are scalable in both size and production volume by asymmetrizing a commercial trekking mattress through duct tape application over its soldering lines. A key advantage of this approach is that actuation is achieved by adhesion of a lightweight, potentially temporary and reusable, skin-like patch to an existing structure, requiring no modification to the structure itself.

\newpage
\section{Conclusion}

The contribution of the present study is twofold. First, we have provided quantitative models describing the pressure-dependent geometry and rigidity of inflatable tubes made of thin sheets, highlighting the role of the dimensionless pressure $P^* = PW^3/B_+$ and the stiffness ratio $B_+/B_-$ for asymmetric tubes. . Second, we have introduced a class of inflatables with broad shape-programming capabilities as soft actuators. The large range of available curvatures is governed by the coupling between geometric bonding of sheets and their mechanical response to pressurization, with self-contact setting both a limit on curvature and a mechanism for stiffening.

Our results challenge the paradigm of geometric volume maximization in the design of thin-sheet inflatables -- which restricts descriptions to the large $P^*$ limit while neglecting the influence of pressurization and sheet stiffness on structural rigidity. In contrast, our model operates throughout inflation, enabling continuously programmable devices with precise curvature control across a large range of pressures.

We close on the question of biological analogy, often referred to as bioinspiration. The comparison between inflatables and pressurized plant cells is a recurring motif in the literature of inflatables~\cite{dessauce1999inflatable}, and stiffness asymmetry is known to drive curvature in plant tissues. A recent study by Chandler~\textit{et al.} pursued this analogy directly, fusing chains of inflatables pouches as a model for plant cells, accounting for both bending and stretching in the sheets~\cite{chandler2025mechanics}. Their geometry, however, differs in a key way: the presence of horizontal walls between cells pins the junction angle externally and suppresses the boundary-layer effects central to the shape-morphing behavior reported here. The two systems are mechanically distinct, and our study appears to have little relevance to plant mechanics due to sharp cusps at junctions. The proposed asymmetric tubes may actually bear a closer resemblance to insect wings~\cite{rajabi2022insect}.

Of more immediate interest, the proposed thin-sheet tube networks actually offer properties not readily found in biological matter: a lightweight, heat-sealable construction that can be added onto existing structures, combined with a pressure-tunable stiffness, makes them particularly well-suited to soft robotics and deployable structures.

% Experimental section

\section{Experimental Section}

\paragraph{Fabrication.} Networks of tubes are made by heat-soldering two fusable sheets of TPU-coated nylon (ExtremTextil) -- the setup is further detailed in Supplementary section B. To make tubes with low stiffness asymmetry ($B_+/B_- < 15$), we solder two sheets differing in stiffness. To make tubes with larger stiffness asymmetry, we solder sheets of identical stiffness and then glue plastic sheets on one side using a neoprene-based glue (Bostik 1400). To measure the junction angle for a single tube (Fig.~\ref{fig:cross-section}a-d, we fabricate single tubes with soldered airtight fins along their edges as visual markers, as shown in Fig.~\ref{fig:cross-section}a. We considered slender tubes only, with a length at least 10 times larger than their width. Connection between neighboring tubes were designed as rounded and separate enough to minimize their impact on the geometry of tubes.\\

\paragraph{Bending stiffness of individual sheets.} The bending stiffnesses of the nylon sheets are determined through the measurement of the natural frequency of a ribbon made of the material~\cite{vani2025stiffness}. We measure the oscillations of the ribbon using a laser sensor (Micro-Epsilon optoNCDT 1900). A typical bending stiffness per unit length of a sheet of weight $220\,\mathrm{g/m}^2$ is $4\,$mN.m. \\

\paragraph{Inflation set-up.} We inflate networks of tubes using the pressurized air supply from the building.
The pressure is set by a pressure controller (SMC IR200000) and measured with an electronic sensor (Adafruit MPRLS).\\

\paragraph{Transverse stiffness.} Two tubes are set in a traction machine (Instron 6865), held at each extremity through floppy fins which transmit forces only. The experiment is realized quasi-statically, measuring visually the angle between the axis of the two tubes $\pi - 2\varphi$ and the lever arm $h$.\\

\medskip
\textbf{Supporting Information} \par
Supporting Information is appended at the end of this document.

% Acknowledgements
\medskip
\textbf{Acknowledgements} \par
This work received financial support from Agence Nationale de la Recherche (ANR MatAIRial 22-CE51-0024). N.V. was supported by a doctoral fellowship awarded by ENS Paris-Saclay. The authors are grateful to Vanshika Singhania for her assistance with preliminary experiments. The authors thank Fabian Brau, Basile Audoly, Mélina Skouras, and Arthur Lebée for discussions, as well as Diane Komaroff, Camille Aracheloff, Alex Leguen, Matteo Milani, and Elias Lundheim for technical help. Fabrication of the inflatables was refined through the authors' participation to Automorph's Creative Differences pavilion during the 2023 London Design Biennale, with contributions from Ofri Dar, Arielle Blonder and Noga Zajfman. N.V. thanks Samantha Kucher for noticing the tulips along the way, and the groundskeepers of the Jussieu campus for nurturing them.

\clearpage
\begin{center}
{\Large\bfseries Supporting Information}\\[2mm]
{\large Pressure and asymmetry govern the shape and stiffness of inflatables}\\[2mm]
Nathan Vani$^{1,2,\ast}$, Tom Joblin$^{1}$, Alejandro Ibarra$^{1,3,4}$, Jos\'e Bico$^{1}$, \'Etienne Reyssat$^{1}$, Beno\^it Roman$^{1}$\\[2mm]
\small
$^{1}$Physique et M\'ecanique des Milieux H\'et\'erog\`enes, CNRS, ESPCI Paris, Universit\'e PSL, Sorbonne Universit\'e, Universit\'e Paris Cit\'e, 7 quai Saint-Bernard, 75005 Paris, France\\
$^{2}$Institute of Physics, University of Amsterdam, Amsterdam, the Netherlands\\
$^{3}$Experimental Soft Matter Physics Group, Department of Physics and Materials Science, University of Luxembourg, Luxembourg L-1511, Luxembourg\\
$^{4}$Institute for Advanced Studies, University of Luxembourg, Belval Esch-sur-Alzette, Luxembourg\\
$^\ast$nathan.vani@ens-paris-saclay.fr
\end{center}
\vspace{4mm}

This document contains descriptions of the supplementary videos (section A), experimental methods (section B), cross-section model with additional experimental and analytical results (section C), limits of the model for high pressures (section D), analytical and numerical modeling of the post-contact regime (section E), characterization of bending stiffness in tube networks (section F), as well as additional examples of applications (section G) and algorithms for pattern generation and numerical solving of the cross-section model (section H).

\vspace{5mm}

%%%%%%%%%%%%%%%%%%%%%%%%%%%% SUPP. VIDEOS %%%%%%%%%%%%%%%%%%%%%%%%%%%% 
\noindent {\Large {\textsc{A. Supplementary videos}}}

\vspace{3mm}
\noindent We provide several videos illustrating the behavior of the system.

\begin{itemize}
\item[(1)] MovieS1.mp4 -- manual inflations of an asymmetic inflatable ($B_+/B_- = 25$, $W=1$ cm, maximum pressure of the order of 1 bar).
\item[(2)] MovieS2.mp4 -- actuation of a flower fitted with a network of tubes
\item[(3)] MovieS3.mp4 -- numerical resolution of the inflation of a cross-section

\end{itemize}

\clearpage

%%%%%%%%%%%%%%%%%%%%%%%%%%%% fabrication %%%%%%%%%%%%%%%%%%%%%%%%%%%% 

\noindent {\Large {\textsc{B. Fabrication methods}}}

Airtight network of inflatable channels are fabricated by heat-sealing two planar sheets. Besides the novel use of asymmetry in thin-sheets inflatables, we have reduced manufacture time and improved soldering quality with the use of a carbon-coated Teflon sheet compared to previous iterations of the method~\cite{niiyama2015pouch, ou2016aeromorph, siefert2019programming, siefert2020programming}.

\begin{figure}[h!]
    \centering
    \includegraphics[width = 0.8\textwidth]{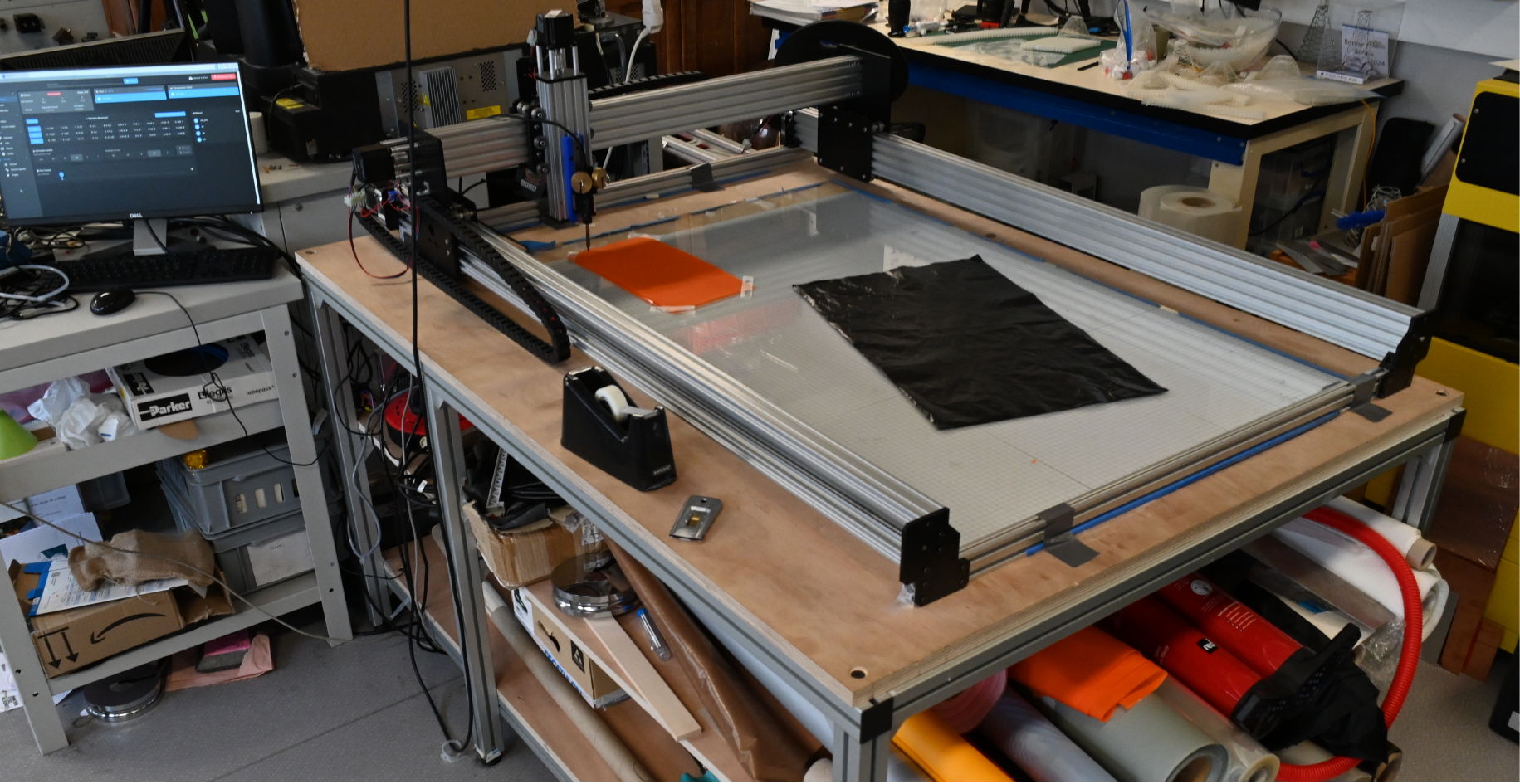}
    \caption{Fabrication setup. The table, made of Norcan profiles and wood plates, holds the plotter machine set above a glass plate. The motion of the arm controls the soldering iron which points downward towards the plate.}
    \label{outils:table}
\end{figure}

\paragraph*{Fabrication set up.} Fig.~\ref{outils:table} shows the setup used to make inflatables. A three-axis arm (OpenBuilds WorkBee, referred to as a computer numerical control (CNC)) is mounted with a soldering iron through a 3D-printed adaptor. The iron is commanded in temperature by a soldering station (Weller WE 1010 70W) and fitted with a rounded conical tip (typically a Weller WEETA $1.6\,$mm). The motion of the arm is controlled by a 3D-printer controller (Duet3D) which comes with its own G-code interpreter software. We set a constant iron velocity $v$ for soldering of the order of \SI{100}{\milli\meter/\minute}. The pressure applied on the textiles is controlled by mounting the iron on a freely sliding rail along the vertical direction. This ensures that the applied force depends only on the weight of the soldering iron and attached fixtures, independently of the force applied by the motor of the plotter motor. We attach a clamp on the iron and add additional calibrated weights for a total force applied $mg$ with $g$ the gravitational constant and $m$ the total mass (iron + clamp + weights) of typically \SI{3}{\newton}.

\begin{figure}[h!]
    \centering
    \includegraphics[width = 0.8\textwidth]{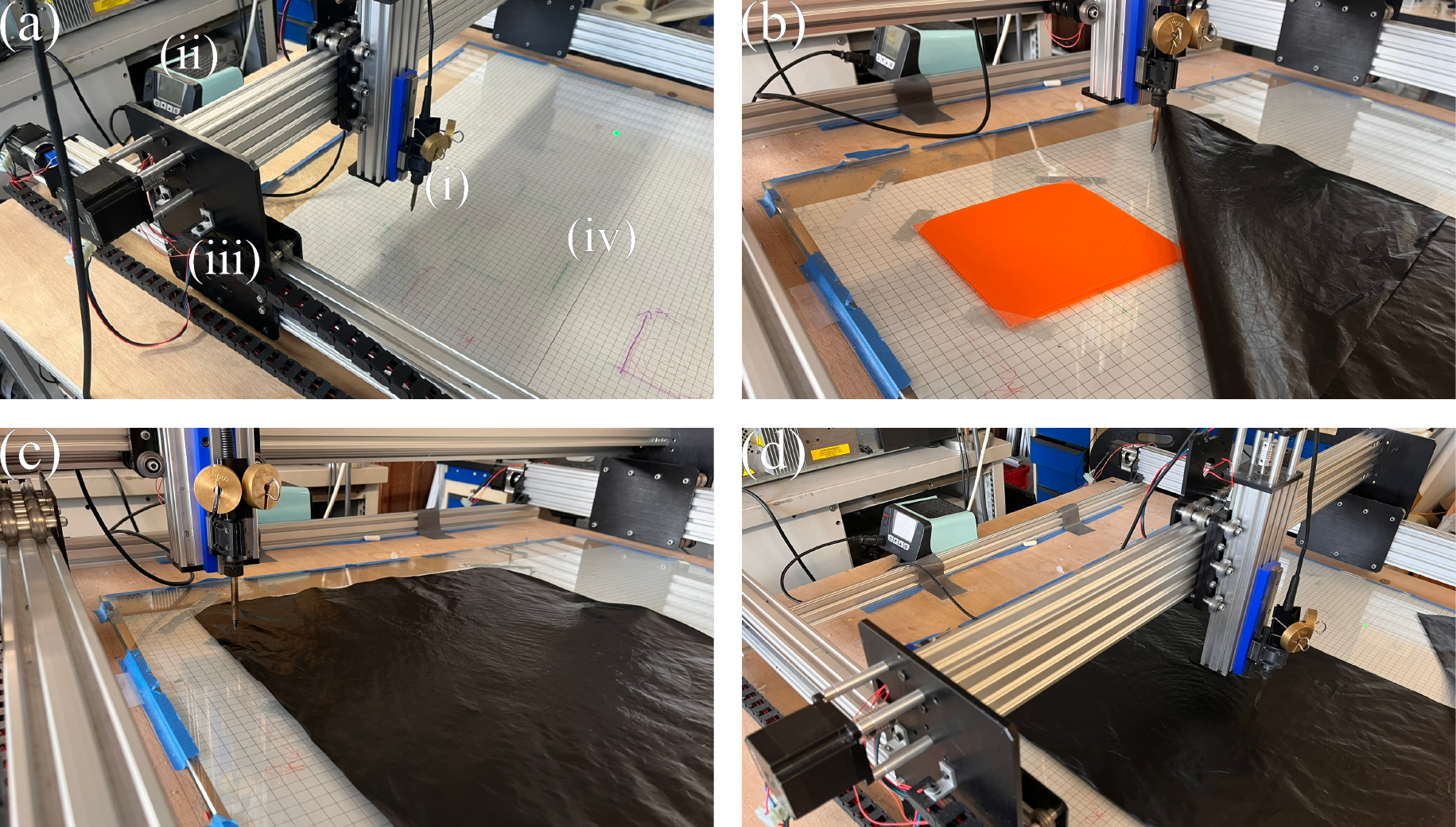}
    \caption{
    Soldering process. (a) Close up of the soldering iron (i) with attached weights and connected to the soldering station (ii). A custom-made support is used to attach the iron to the machine (iii) -- a vertical freely sliding rail connects the iron to the support. A tempered glass plate (iv) is used as a base for the soldering. (b) We first tape two pieces of TPU-coated textiles to the base, setting as well the zeros of the plotter in the three directions. (c) We then cover the textile with a larger sheet of Teflon which we tape independently to the glass plate. (iv) We finally turn on the soldering station to the desired temperature and start the G-code interpreter.}
    \label{outils:quatre_cnc}
\end{figure}

The thin sheets used are woven Nylon textiles (from ExtremTextil) coated on one or both sides with a thermoplastic polyurethane (TPU). The global thickness of the fabrics ranges from 100 to \SI{400}{\micro\meter}, and they have a natural curvature as they are packaged in rolls.
We are also able to solder homogeneous plastic sheets, as long as they are fusible, as in~\cite{vani2025asymmetric}, with polyethylene (PE) and polyethylene terephthalate (PET) sheets. Regardless of the material, the two sheets to be soldered are taped flat on top of each other on the glass plate underneath the plotter.

An important technical improvement is obtained by using a protective layer on top of the textile sheets. We used carbon-coated Teflon sheets (TVTAS080 from Plastiques Elastom\`eres). The protective layer is anti-adhesive and is not damaged by high temperatures ($T >$ \SI{300}{\degreeCelsius}). Its high thermal conductivity leads to better diffusion of the heat, resulting in good seam lines while not damaging the Nylon layer. We can thus use higher temperatures without melting the Nylon, leading to higher soldering velocities. A typical set of parameters is $v =$ \SI{300}{\milli\meter/\minute}, $T =$ \SI{450}{\degreeCelsius} and $m =$ \SI{300}{\gram}.

\paragraph*{Design of patterns.} The process of designing a pattern is made of two steps. We first generate the 2D soldering patterns using Python codes with a parametric characterization or we directly draw the patterns using Inkscape. The patterns need to be generated in a vector-based image format -- here we used Scalable Vector Graphics (.SVG). A code used to generate the pattern for an array of parallel tubes is provided at the end of this document. Once we have created an image, we need to convert it into a series of instructions readable by the machine. The format used to this end is G-CODE. We did this conversion using the proprietary software provided by OpenBuilds, the maker of the plotter.

\paragraph*{Asymmetric inflatables.} Two techniques are used to fabricate stiffness asymmetric tubes. In the most straightforward one, we weld two sheets differing in stiffness. This creates an object with a uniform asymmetry in stiffness. While straightforward, this method is nevertheless limited to low values of stiffness asymmetry, as the two materials must be fusible. With commercially available textiles, we could not make inflatables with a stiffness asymmetry $B_+/B_-$ larger than $20$, nor with very large stiffnesses ($B_+ < \SI{10}{\milli\newton\meter}$). It is nonetheless a reliable and simple method of fabrication, allowing for a good control.

In another method, we weld two identical sheets, creating a symmetric network of tubes which we then render asymmetric by gluing on one of its sides a patch of stiffening material. This allows to reach higher values of stiffness asymmetries. The glue used is neoprene-based (Bostik 1400), and various sheets can be used as stiffening patches, most notably PET (Mylar brand, 50 to \SI{250}{\micro\meter} thick) and PLA-coated paper (\SI{300}{\micro\meter} to \SI{1.5}{\milli\meter}). However, thin steel sheets can also be used -- or any material which can be glued and sustain the strains and stresses associated with inflation.

Experiments in this paper were performed with inflatables corresponding to five different stiffness asymmetries, corresponding to five different couples of materials. Two couples have a low stiffness asymmetry ($B_+/B_- = 8$ and $10$) and are obtained through the welding of distinct textiles sheets (respectively of 90 and \SI{160}{\micro\meter}
%160 micron 
thick sheets, 
and 160 and \SI{250}{\micro\meter}
%250 microns 
thick sheets). Three other couples have a larger stiffness asymmetry ($B_+/B_- = 25$, $50$ and $800$) and were obtained through the second method with tubes made of two identical sheets (
\SI{160}{\micro\meter}
%$160$ micron 
thick Nylon), gluing respectively $100$, 
\SI{200}{\micro\meter}
%$200$ microns 
thick PET and 
\SI{750}{\micro\meter}
%$750$ microns 
thick PLA.

It is interesting to note that the Nylon sheets are anisotropic with regard to their bending stiffness. It is thus possible to fabricate asymmetric inflatables from identical sheets by superimposing them along crossed orientations before welding. With typical Nylon sheets, crossing orientations can lead to stiffness asymmetries up to $B_+/B_- = 5$ -- though it is difficult to control precisely the relative orientation. We emphasize here that we still had to be particularly careful about the orientation of the sheets for the experiments as their orientation can have a large impact on the ratios of asymmetry -- especially when one is aiming to fabricate symmetric networks.

\paragraph*{Inflation.} Once the inflatable is fabricated, the newly created inflatable are cut out from the surrounding textile. Microfluidics connectors made of polypropylene with luer locks are then fitted in the opening. These are standard components, available from Amazon or Darwin Microfluidics, and we use connectors of either size 3/32 or 1/8 inches. The channel opening is precisely designed to be small enough to ensure a tight fit of the fluidic connector, resulting in a mostly airtight connection. Superglue can be applied around the connector to improve airtightness.

\begin{figure}[h!]
    \centering
    \includegraphics[width = 1\textwidth]{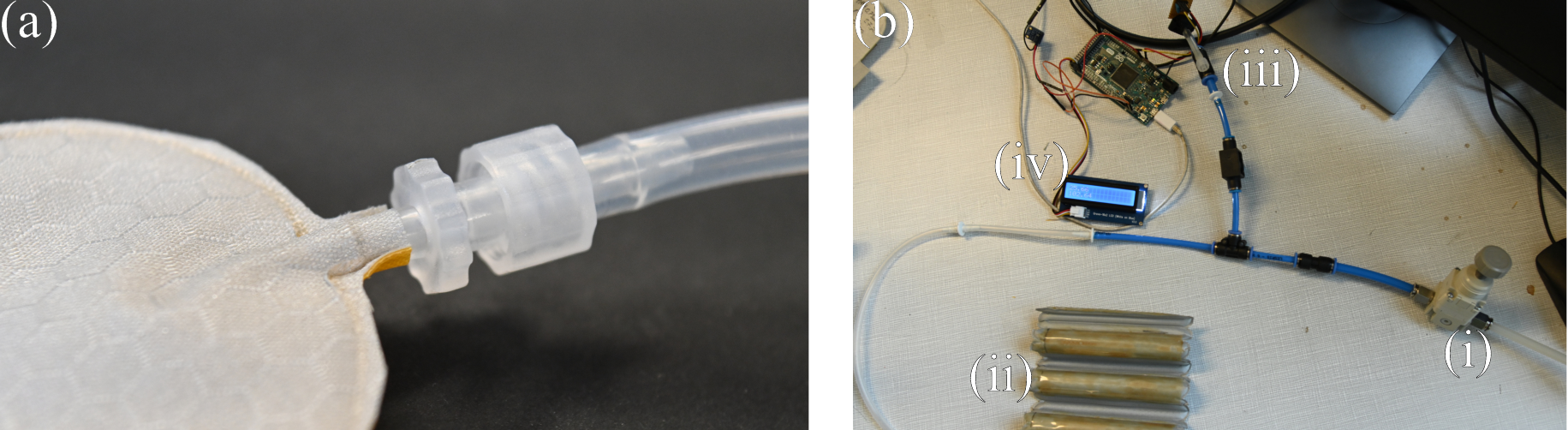}
    \caption{(a) Close-up of a fitted in connector allowing for pressurization of a tube. (b) Setup to control input pressure with a controller (i) of an inflatable (ii) while a sensor (iii) measures pressure and displays it on a LED screen (iv).}
\label{outils:osci_graph}
\end{figure}

The connector can then be attached through its male or female counterpart to a silicone tube for pressurization. Quantitative experiments were performed using the pressurized air network from the building, and the pressure was set by a pressure controller (SMC IR20000) and measured by an electronic sensor (Adafruit MPRLS). Some of the qualitative experiments were done with a manually activated bulb pump.

\clearpage

%%%%%%%%%%%%%%%%%%%%%%%%%%%% Model cross-section %%%%%%%%%%%%%%%%%%%%%%%%%%%% 
\noindent {\Large {\textsc{C. Model for the cross-section}}}

\vspace{5mm}

\begin{figure}[h!]
    \centering
    \includegraphics[width = 0.7\textwidth]{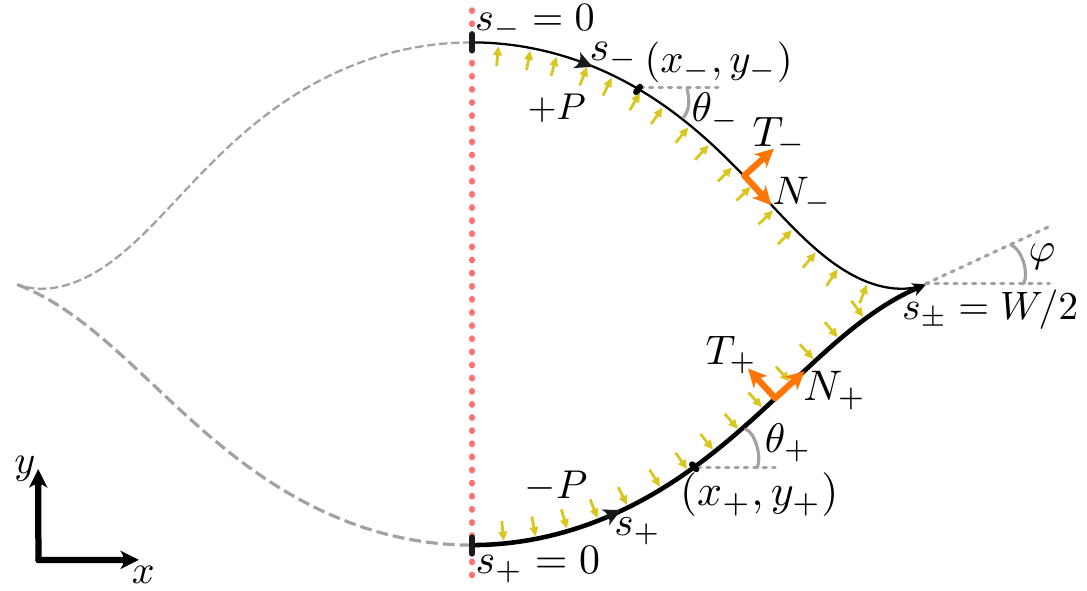}
    \caption{Sketch of the coupled rods modeling the cross-section. Note that we count angles positively counterclockwise, the drawn angles $\theta_+$ and $\theta_-$ are thus respectively positive and negative.}
    \label{tubes:kirchhoff}
\end{figure}

\paragraph{Presentation of the system.} From the network of tubes, we reduce the object of study to a single asymmetric tube. The geometry of slender straight tubes is invariant by translation along their length. This holds true sufficiently far from the two ends of the tube. With $W$ the width of the tube while deflated and $L$ its length, we restrict our study to tubes with a ratio of $L/W > 10$. By staying in this regime, we reduce the study to the static analysis of the cross-section of a single tube. We describe thereafter this model as presented in Fig.~\ref{tubes:kirchhoff}.

Each sheet making up the tube is represented by a Kirchhoff rod. Here, the assumptions of inextensibility and linear elasticity mostly amount to the tubes being in a deformation regime lower than the yielding threshold. The bending stiffnesses considered, $B_+$ and $B_-$ correspond to the stiffness per unit of length of the sheets making up the tubes. We introduce for each rod, noted with $+$ and $-$ subscripts, their arc lengths as $s_\pm$, their local angles and local positions in the reference frame as $\theta_\pm$ and $(x_\pm, y_\pm)$, the internal forces normal to their section $N_\pm$ and the forces transverse to the rod $T_\pm$. The curvatures $\kappa_\pm$ correspond to the first derivative of the local angles with respect to their respective arc lengths. The local torques correspond to $B_\pm\kappa_\pm$ following the assumption of linear elasticity.

Each variable can be made dimensionless as such:

\begin{equation}
    \begin{cases}
        N^*_\pm = \frac{W^2}{B_+}N_\pm, \\
        T^*_\pm = \frac{W^2}{B_+}T_\pm, \\
        \kappa^*_\pm = \kappa_\pm W, \\
        s^*_\pm = s_\pm /W, \\
        (x^*_\pm, y^*_\pm) = (x_\pm /W, y_\pm /W).
    \end{cases}
\end{equation}

The force and torque balance equations are:

\begin{equation}
    \begin{cases}
        \dfrac{\dd \kappa_+^*}{\dd s_+^*} = - T_+^* \\[8pt]
        \dfrac{\dd N_+^*}{\dd s_+^*} = T_+^*\kappa_+^* \\[8pt]
        \dfrac{\dd T_+^*}{\dd s_+^*} = - N_+^*\kappa_+^* + P^*    
    \end{cases}
\label{tube_kirchhoff_plus_adim}
\end{equation}

\begin{equation}
    \begin{cases}
        \dfrac{\dd \kappa_-^*}{\dd s_-^*} = - \frac{B_+}{B_-}T_-^* \\[8pt]
        \dfrac{\dd N_-^*}{\dd s_-^*} = T_-^*\kappa_-^* \\[8pt]
        \dfrac{\dd T_-^*}{\dd s_-^*} = - N_-^*\kappa_-^* - P^* 
    \end{cases}
\label{tube_kirchhoff_moins_adim}
\end{equation}

\noindent differing in the orientation of the pressure only. The six kinematic equations are:

\begin{equation}
    \begin{cases}
        \dfrac{\dd \theta_\pm}{\dd s_\pm^*} = \kappa_\pm^*, \\[8pt]
        \dfrac{\dd x_\pm^*}{\dd s_\pm^*} = \cos \theta_\pm, \\[8pt]
        \dfrac{\dd y_\pm^*}{\dd s_\pm^*} = \sin \theta_\pm. \\[8pt]
    \end{cases}
\end{equation}

Let us note here that since the cross-section exhibits a reflective symmetry about the middle point of the section, we solve the system only over half of the cross-section. We orient the arc lengths from left to right, starting at $s_\pm = 0$ at the middle point and ending both at $s_\pm = W/2$ at the junction of the two rods.

To solve this system of twelve ODEs, twelve boundary conditions are needed. Along the symmetry line, we impose vanishing transverse forces $T_\pm$ as well as zero slopes -- both results stemming from the left-right symmetry of the system. The latter condition makes the system stable in rotation. We also set the position of one point in the system, arbitrarily choosing the one corresponding to $s_+ = 0$. The conditions at the center of the cross-section are:

\begin{equation}
    \begin{cases}
        T^*_+(s_+^*=0) = 0, \\[8pt]
        T^*_-(s_-^*=0) = 0, \\[8pt]
        \theta_+(s_+^*=0) = 0, \\[8pt]
        \theta_-(s_-^*=0) = 0, \\[8pt]
        x_+^*(s_+^*=0) = 0, \\[8pt]
        y_+^*(s_+^*=0) = 0.
    \end{cases}
\label{tube:boundary_condition_midpoint}
\end{equation}

At the junction point, the conditions correspond to the equality of torque and forces, as well as the matching of the positions and angles:

\begin{equation}
    \begin{cases}
        \kappa^*_+(s_+^*=1/2) + \frac{B_-}{B_+}\kappa^*_-(s^*_-=1/2) = 0, \\[8pt]
        N^*_+(s_+^*=1/2) + N^*_-(s^*_-=1/2) = 0, \\[8pt]
        T^*_+(s_+^*=1/2) + T^*_-(s^*_-=1/2) = 0, \\[8pt]
        x^*_+(s^*_+=1/2) = x^*_-(s^*_-=1/2), \\[8pt]
        y^*_+(s^*_+=1/2) = y^*_-(s^*_-=1/2), \\[8pt]
        \theta_+(s^*_+=1/2) = \theta_-(s^*_-=1/2).
    \end{cases}
\label{tube:boundary_condition_junction}
\end{equation}

The junction angle $\varphi$ as discussed in the main paper corresponds here to $\theta_\pm(s_\pm=W/2)$ as shown in Fig.~\ref{tubes:kirchhoff}. We note that other conditions could be used in this system, for example the equality of $x_+(s_+=0)$ and $x_-(s_-=0)$ is another way of imposing torque balance in the system.

The system of 12 dimensionless ODEs is a well-posed boundary-value problem. We solve the system with the Python package SciPy using the solve\_bvp method.

For maximal stability, we solve the system in the following way. First, we set a ratio of stiffness asymmetry $B_+/B_- \geq 1$ and an initial low dimensionless pressure $P^*$, typically $10^{-2}$. We set an initial value to the 12 variables of the problem chosen based on the linear solution to the problem: $x_\pm^*$ are set as $s_\pm^*$, $y_\pm^*$ and $N_\pm^*$ as zero, $\theta_\pm$ as a linear function of the pressure and $s_\pm^*$ to the cube, $\kappa_\pm^*$ as a linear function of the pressure and $s_\pm^*$ squared, $T_\pm^*$ as a linear function of the pressure and $s_\pm^*$. We then solve the boundary-value problem using a continuation method: we use the previous solution as an initial guess for each step while slowly increasing the pressure. We typically consider rods made of 100 nods and around 1000 logarithmically increasing steps for the pressure.

\begin{figure}[h!]
    \centering
    \includegraphics[width = 0.7\textwidth]{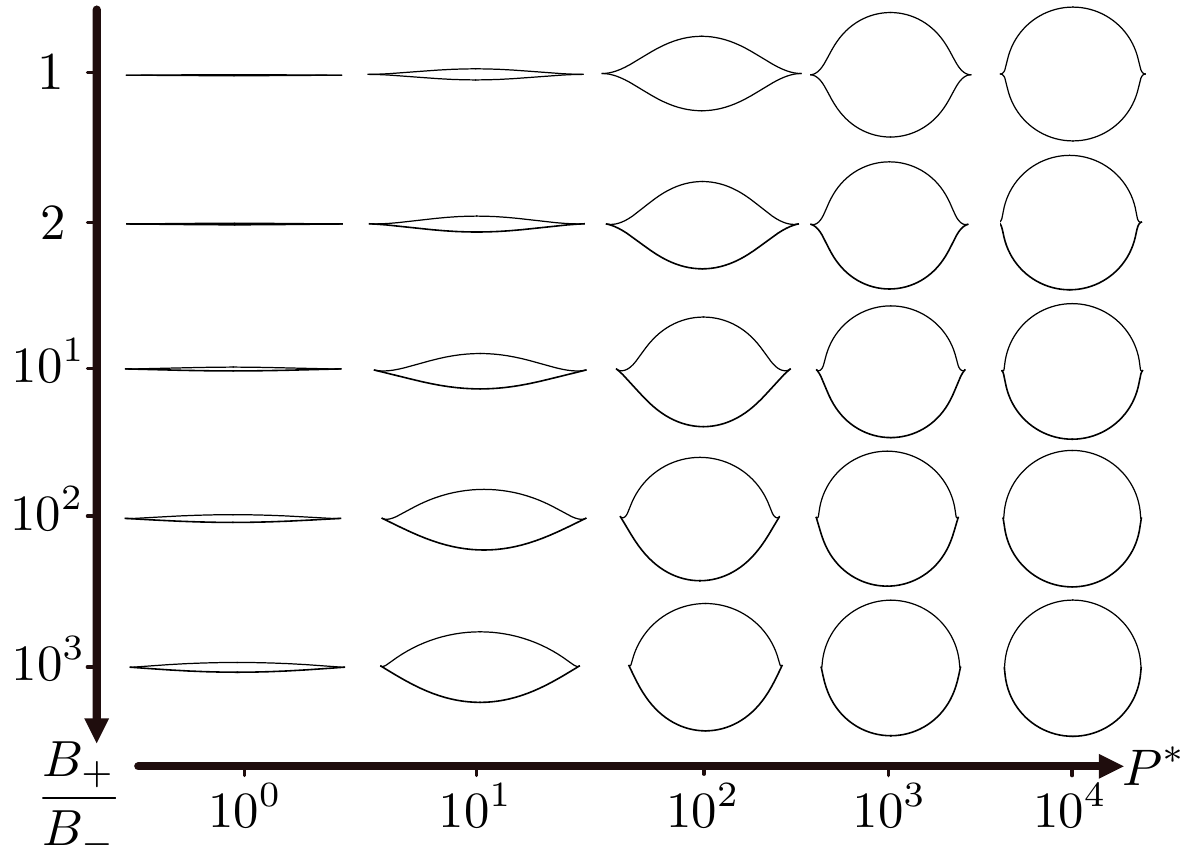}
    \caption{Various cross-sections obtained through numerical resolution of the coupled Kirchhoff rod models for various stiffness asymmetries (vertical axis) and various dimensionless pressures (horizontal axis).}
    \label{tubes:ronds}
\end{figure}

\begin{figure}[h!]
    \centering
    \includegraphics[width = 0.85\textwidth]{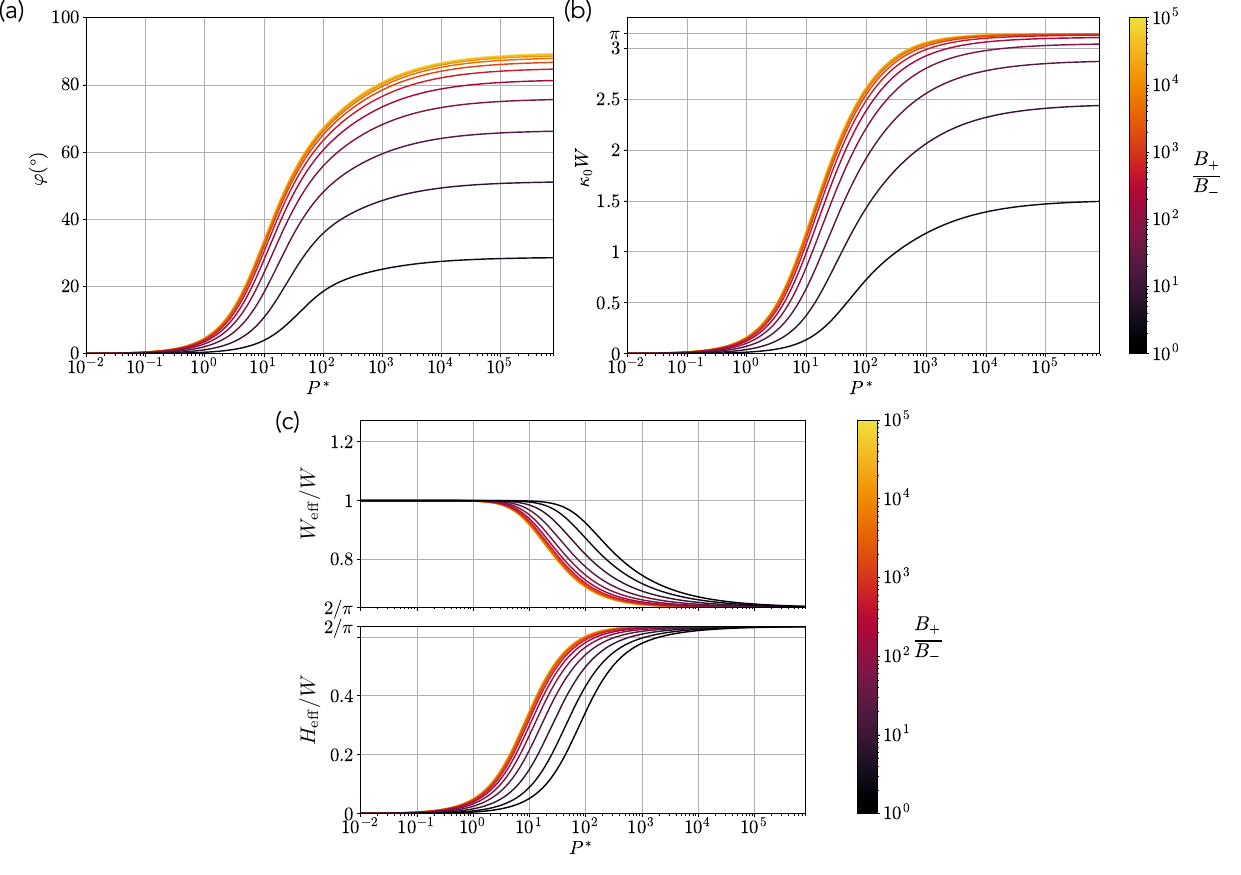}
    \caption{Evolution of geometrical parameters as a function of $P^*$ as predicted from the Kirchhoff rod models: (a) junction angle $\varphi$, (b) dimensionless curvature $\kappa_0W$, (c) effective width $W_{\mathrm{eff}}/W$ and height $H_{\mathrm{eff}}/W$.}
    \label{SM:results_kirchhoff}
\end{figure}

\paragraph*{Numerical predictions.} Fig.~\ref{tubes:ronds} shows cross-sections for increasing dimensionless pressure and stiffness asymmetry. Fig.~\ref{SM:results_kirchhoff} shows the evolution of geometrical parameters as a function of $P^*$ of the simulations for various ratio of stiffness asymmetry.

\begin{figure}[h!]
    \centering
    \includegraphics[width = \textwidth]{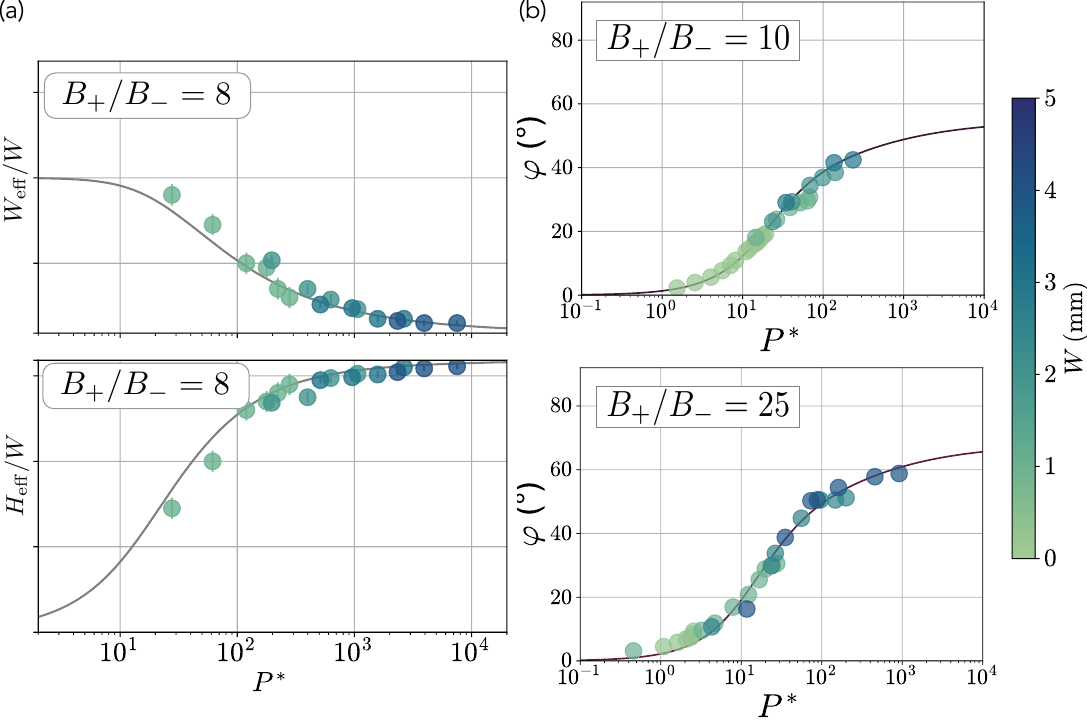}
    \caption{Evolution of the geometrical parameters as a function of $P^*$ according to experiments (markers) for various tube widths (colorbar), compared to predictions from cross-section model (lines) : (a)$H_{\mathrm{eff}}$ and $W_{\mathrm{eff}}$ for $B_+/B_- = 8$ ; (b)$\varphi$ for $B_+/B_- = 10$ and $25$.}
    \label{SM:add_data_geo}
\end{figure}

\paragraph*{Additional data.} Complementary experiments to the main paper are plotted in Fig.~\ref{SM:add_data_geo}, showing the evolution of the effective width and height for an asymmetric tube as well as the junction angle for two additional ratios of asymmetry. We find a proper agreement in the slender ($\ell < W$) regime as well.

\clearpage

\paragraph*{Asymptotic regime of low pressure and large asymmetry.} We propose here an energetically method to predict the junction angle in the double asymptotic regime of small pressures and alrge stiffness contrast.

\begin{figure}[h!]%% placement specifier
    \centering%% For centre alignment of image.
    \includegraphics[width=0.8\linewidth]{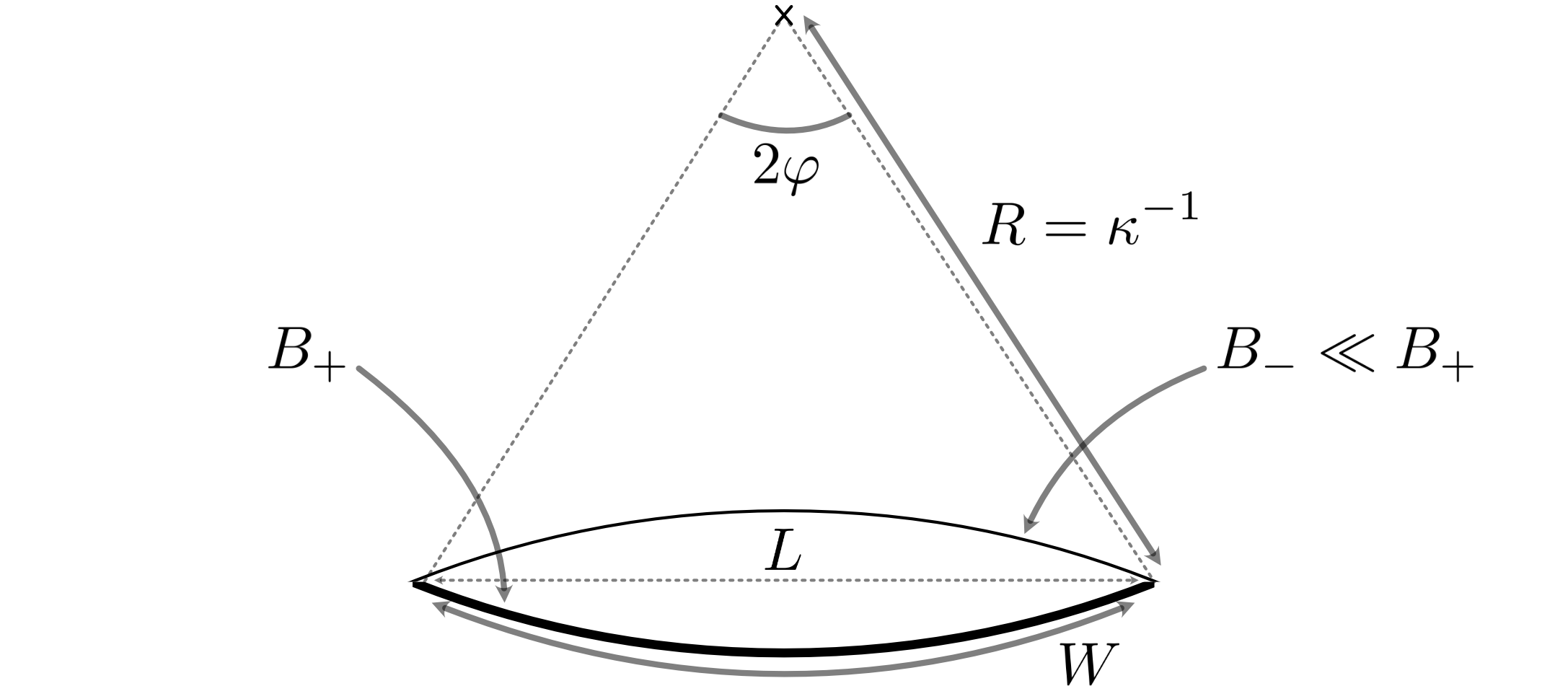}
    \caption{Sketch of the system idealized in the asymptotic regime of vanishing weaker stiffness.}
    \label{fig:schema_scaling}
\end{figure}

We consider the asymptotic case of one rod of stiffness $B_+$ connected to a membrane of vanishing bending stiffness ($B_- \simeq 0$).  We note $P$ the applied pressure, $W$ the length of this rod, $\kappa = R^{-1}$ its curvature, and $\varphi$ the associated angle as drawn in Fig.~\ref{fig:schema_scaling}. The balance of the bending energy, stored entirely in the stiff rod, and the work of pressure is used to estimate the junction angle. The bending energy of the rod is:
\begin{equation}
 U_{bending}=\frac{1}{2}B_+\kappa^2W   
\end{equation}
or, noting that $\kappa=2\varphi/W$,
\begin{equation}
    U_{bending}=2\frac{B_+ \varphi^2}{W}    
\end{equation}
The work done to inflate the volume $V$ of the tube at pressure $P$ is:
\begin{equation}
    U_{pressure}=PV
\end{equation}
Considering that the walls of the tube are two arcs of circles of radius $R$, the volume (per unit width) of the tube is:
\begin{equation}
    V=R^2\left(\varphi-\cos \varphi \sin \varphi \right)
\end{equation}
or, in the limit of small angles and given that $R=W/(2\varphi)$:
\begin{equation}
    V \simeq \frac{W^2\varphi}{6}
\end{equation}
Balancing the work of pressure and the bending energy then yields:
\begin{equation}
    \varphi \simeq \frac{PW^3}{12B_+}=\frac{P^*}{12}
\end{equation}
This minimal model gives the numerical coefficient $1/12$ 
in excellent agreement with the slope determined numerically with the complete model (yellow line for large $B_+/B_-$ in figure 2g(iii) of the main text).

\paragraph*{Approximate values.} Finally, we provide empirical formulas for the geometrical parameters. In the limit of low pressure, the response of the system in terms of junction angle, effective width and curvature is linear. The most important parameter in this regime is the curvature, especially for particularly asymmetric networks. The numerical data is found to collapse well with the following expression for $P^* < 10$:

\begin{equation}
    \kappa_0W = \frac{1}{6}P^* \exp\left(\frac{-5}{\sqrt{B_+/B_-}}\right).
    \label{empi1}
\end{equation}

\begin{figure}[h!]
    \centering
    \includegraphics[width = \textwidth]{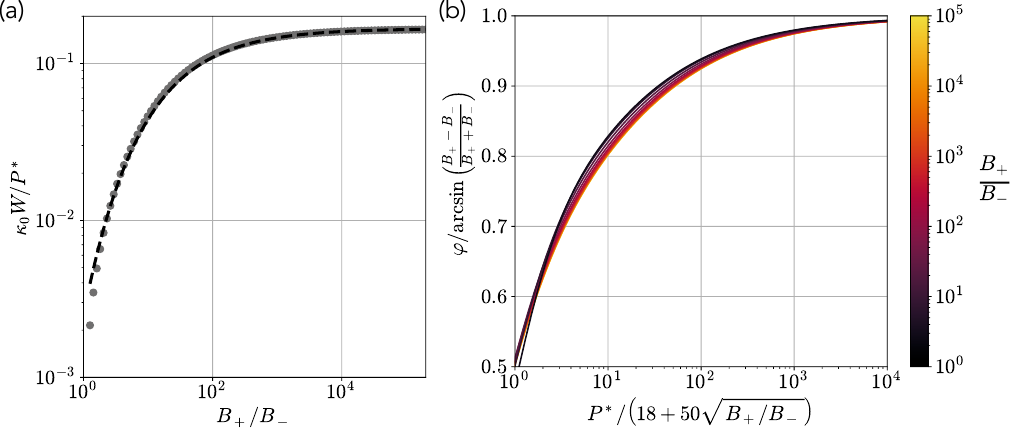}
    \caption{(a) Slope of the evolution of the curvature $\kappa_0$ in the linear regime at low pressures as predicted by the numerical model (markers) and through the empirical Eq.~\eqref{empi1}. (b) Collapse of the numerical data for the junction angle using Eq.\eqref{empi2}.}
    \label{SM:empifig}
\end{figure}

\noindent
The slope predicted by the expression is plotted along numerical predictions in Fig.~\ref{SM:empifig} (a). We further provide an empirical fit of $\varphi$ as a function of the dimensionless pressure:

\begin{equation}
    \varphi =\arcsin\left(\frac{B_+ - B_-}{B_+ + B_-}\right) \frac{P^*}{18 + 50\sqrt{B_+/B_-}}.
    \label{empi2}
\end{equation}

\noindent
The partial collapse of the numerical predictions using this expression is shown in Fig.~\ref{SM:empifig} (b). Note that Eq.~\eqref{empi1} is not physical in the $B_+/B_- \to 1$ limit and Eq.~\eqref{empi1} is not physical in the $P^* \to 0$ limit: both fits are only approximations for designers to get a better intuition over the dependency of the bending on pressure and asymmetry.

\clearpage

%%%%%%%%%%%%%%%%%%%%%%%%%%%% Out of slenderness %%%%%%%%%%%%%%%%%%%%%%%%%%%% 
\noindent {\Large {\textsc{D. Limits of the model for high pressures}}}
\vspace{5mm} 

We discuss the case of high pressures in which the cross-section of the tubes does not follow the assumptions of slenderness and elasticity in the region of the junction. Indeed, in the results on the junction angle for a single tube as represented in Fig. 2 of the main paper, we stopped reporting results when the tubes displayed plastic deformations or a threshold in slenderness was surpassed.

\paragraph{Slenderness assumption.} The largest curvatures are observed in the boundary layer, and the slenderness assumption is broken first
when the typical radius of curvature becomes of the same order of magnitude as the thicknesses of the textiles. The typical sizes $\ell_\pm$ of the boundary layer of each sheet scale as:

\begin{equation}
    \ell_\pm = \sqrt{\frac{\pi B_\pm}{PW}}.
\end{equation}

It is typically considered that the slenderness assumptions hold when there is at least a ratio of 10 between the thickness of the rod and the maximal curvature in the rod~\cite{audoly2000elasticity}. With $t_\pm$ the thicknesses of the sheets, the thresholds are thus $\ell_\pm > t_\pm/10$. The threshold for the inflating pressure is:

\begin{equation}
    P_{slender} = \min \left( \frac{\pi B_+}{1000Wt_+^2}, \frac{\pi B_-}{1000Wt_-^2}\right),
\label{tubes:slender_threshold}
\end{equation}

\noindent with the relevant term being the one related to the weaker membrane in most cases. Indeed, for a beam of Young's modulus $Y$ and Poisson's ratio $\nu$, the ratio of $B/t^2$ reduces to $Yt/(12(1-\nu^2))$. 

\paragraph{Plasticity.} In practice for the considered textiles, the threshold in slenderness is reached before the plasticity threshold but once it is reached, the material quickly yields. Indeed when the rods are no longer slender with regard to their radius of curvature at the junction, the deformation across the section of the tube becomes non-linear and prone to plastic effects.

\begin{figure}[h!]
    \centering
    \includegraphics[width = 1\textwidth]{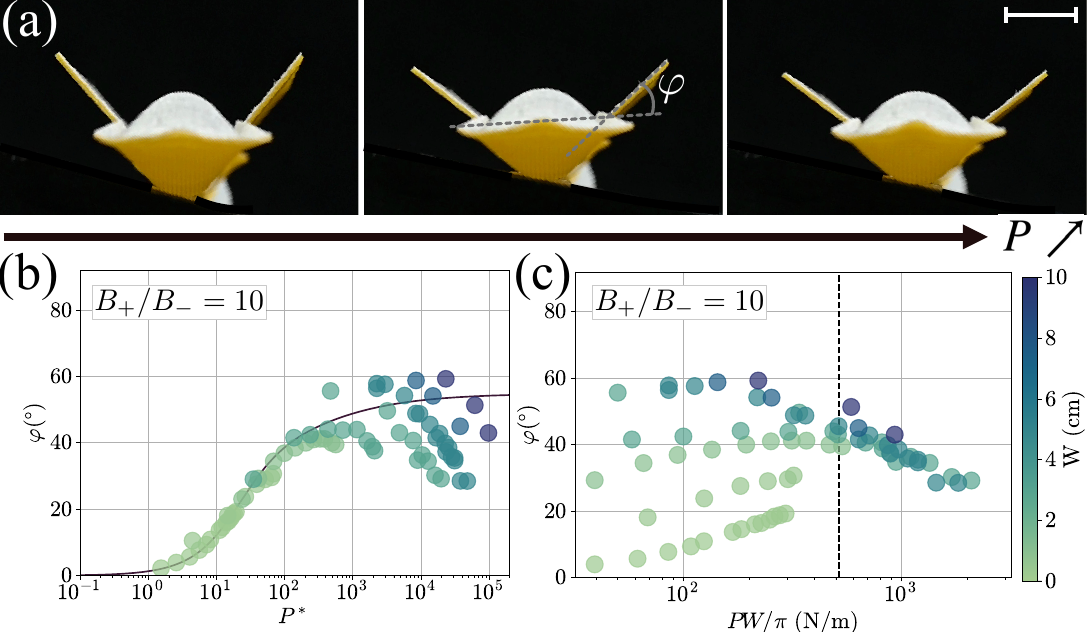}
    \caption{(a) Sequential pictures of a tube of width ($W=$ \SI{2}{\centi\meter}) as the pressure is increased past the assumptions of slender elasticity. The scale bar is \SI{1}{\centi\meter}. (b) Junction angle $\varphi$ as a function of the dimensionless pressure $P^*=PW^3/B_+$. (c) Same junction angle $\varphi$ as a function of the tension in the membrane $PW/\pi$. The dashed line corresponds to the slenderness threshold from Eq.~\ref{tubes:slender_threshold}.}
\label{tubes:out_of_slender}
\end{figure}

Fig.~\ref{tubes:out_of_slender} (a) shows sequential pictures of a tube ($B_+/B_-=10$, $W=$ \SI{2}{\centi\meter}) after reaching the threshold: the junction angle decreases as the pressure increases. We report the junction angles for tubes of varying widths in this regime in Fig.~\ref{tubes:out_of_slender} (b) as a function of the dimensionless pressure $P^*$. First, we note that the trend is general: every tube studied has its junction angle decreasing at large enough pressures. Second, we note that beyond the threshold, the data for different widths $W$ do not collapse anymore. The dimensionless pressure is thus no longer the governing parameter of the system. In the associated experiments, plastic deformation in the weaker sheet near the junction are observed when deflating the tubes.

The tension in the sheets governs the behavior of the sheets in this regime. The tension $N$ in a thin-walled pressurized circular cross-section of diameter $W/\pi$ can be estimated as $N = PW/\pi$. We plot the data as a function of this tension in Fig.~\ref{tubes:out_of_slender} (c). The dashed line corresponds to the threshold roughly estimated by Eq.~\ref{tubes:slender_threshold} and the data collapse  after the threshold is reached. Once the regime of slender elasticity is left, the cross-section still remains circular, while deformations governed by tension rather than bending occur near the junction. These deformations lead to the decrease of the junction angle.

\begin{figure}[h!]
    \centering
    \includegraphics[width = 0.6\textwidth]{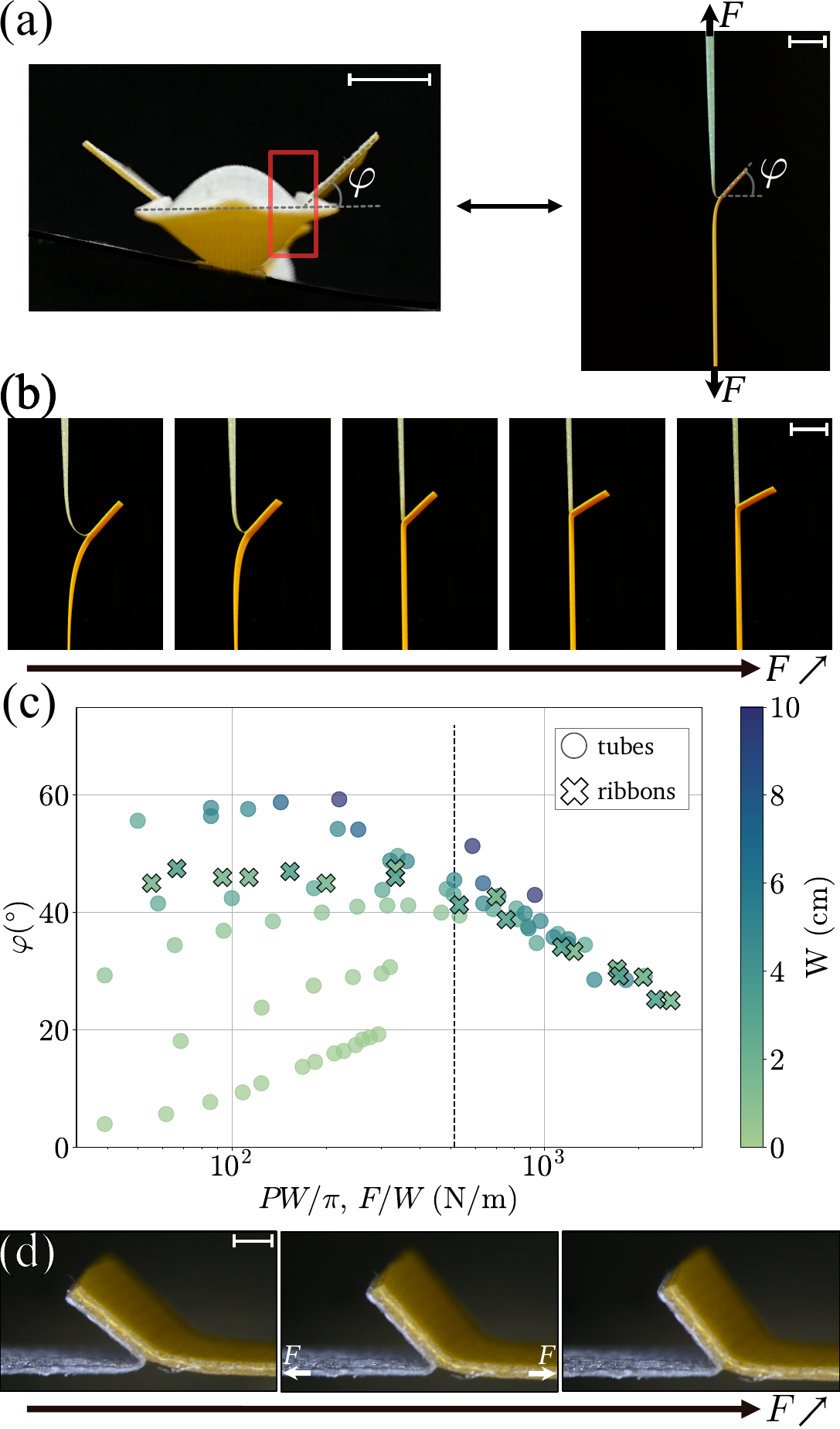}
    \caption{(a) The measurement of the junction angle for a tube at large pressures (left) is analogous to a measurement of the junction angle in the pulling of two bound ribbons (right). Scale bars are \SI{1}{\centi\meter}. (b) Sequential pictures of the junction of two ribbons as the pulling force is increased: the angle is first constant and then starts decreasing after the slenderness assumption is broken. (c) Junction angle for the tube experiments (circles) and ribbon experiments (crosses) as a function of the tensile force per unit of width, respectively $PW/\pi$ and $F/W$ for varying widths $W$ (colorbar). (d) Sequential pictures of the junction with a larger zoom. Scale bar is \SI{0.5}{\centi\meter}.}
\label{tubes:out_of_slender_compa}
\end{figure}

\paragraph{Model experiment.} The decrease of the junction angle observed in Fig.~\ref{tubes:out_of_slender} (c) can be fully captured by an analogous experiment as shown in Fig.~\ref{tubes:out_of_slender_compa} (a) and (b). We weld two ribbons made of the same materials as the tube, and we pull on the ribbons with an increasing force. 
%Results are reported in Fig.~\ref{tubes:out_of_slender_compa} (c). 
The results, shown for ribbons of two different widths $W$, are plotted in Fig.~\ref{tubes:out_of_slender_compa} with the junction angle as a function of the force per unit of length $F/W$. Before the tensile threshold, the junction angle is constant 
%as a function of the force 
as the boundary layer is fully developed -- as discussed in~\cite{vani2025asymmetric}. Beyond the threshold, a decrease of the junction angle is observed, and the data collapse with the one of tubes when plotted as a function of the tensile force $PW/\pi$.

Closer pictures of the junction are shown in Fig.~\ref{tubes:out_of_slender_compa} (d). The mechanisms governing the behavior of the junction for large tensile loads are not trivial. Indeed, several phenomena happen more or less simultaneously. First, since the slenderness assumption is broken, we expect a non-homogeneous distribution of forces in the cross-section. This occurs visibly here concurrently with the yielding.

Since our materials are not homogeneous, being made of woven nylon coated with TPU on one or both sides, the response of the conposite beam is complex. In the experiments reported in Fig~\ref{tubes:out_of_slender_compa}, there is significant stretching happening along the TPU film on both the gray and yellow Nylon. The plasticity itself probably happens throughout the TPU layer. In the case shown here, the curvature becomes so large over a small length that a kink is observed: the deformations localize and lead to a slight fold. We also observe partial debonding of the junction. This area is also particularly difficult to characterize as the soldering process modifies the properties of the two sheets~\cite{broshkevitch2025programmable}. Even predicting the thickness of the bonding layer is challenging.

All of those complexities are material dependent. While we do not provide here any model for the decrease of the junction angle, we can note that this behavior occurs even in the case of two homogeneous elastic ribbons. The trend appears to be general: asymmetric tubes at large enough pressures will see their junction angle decrease. For practical applications of networks of tubes, this phenomenon is thankfully irrelevant as contact between neighboring tubes dominates at large pressures.

\clearpage

%%%%%%%%%%%%%%%%%%%%%%%%%%%% CONTACT %%%%%%%%%%%%%%%%%%%%%%%%%%%% 
\noindent {\Large {\textsc{E. Contact behavior}}}

\vspace{5mm}

\noindent {\large {\textsc{(1). Prediction of contact using the single tube model}}}
\vspace{3mm}

\begin{figure}[h!]
    \centering
    \includegraphics[width = 0.65\textwidth]{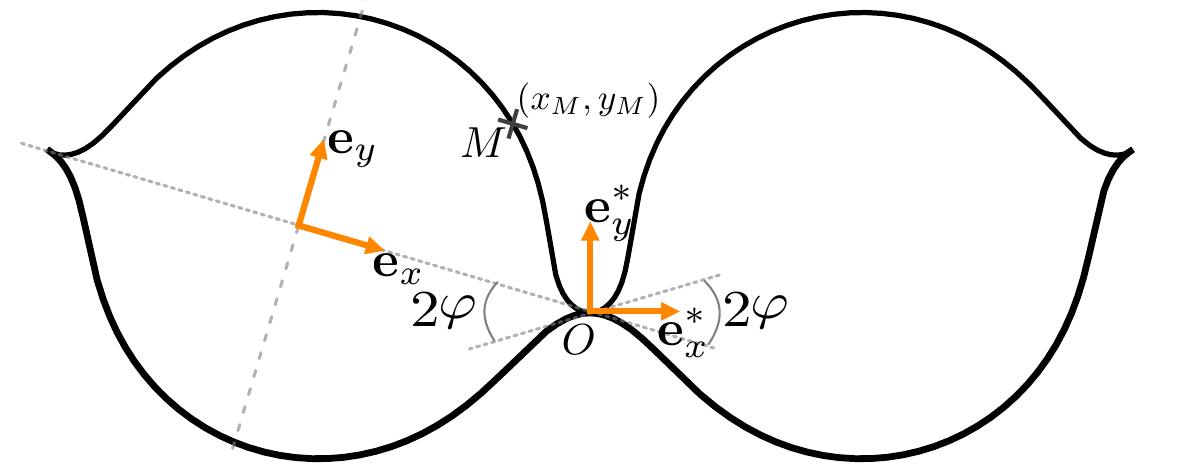}
    \caption{Tubes with two distinct frames of references for the detection of contact.}
\label{tubes:contact_detection_kirch}
\end{figure}

We aim to predict at which dimensionless pressure contact first occurs between two tubes using our coupled rods model as described in section C using the cross-section model. We name this pressure $P_{\mathrm{contact}}$ and the associated junction angle $\varphi_{\mathrm{contact}}$. For two identical neighboring tubes, the system presents an axis of symmetry between the two cross-sections. Fig.~\ref{tubes:contact_detection_kirch} shows two tubes with the frame of reference ($\mathbf{e_x},\mathbf{e_y}$) in which we solve the equations and another one we introduce now as ($\mathbf{e_x^*},\mathbf{e_y^*}$). This frame of reference remains aligned with the seam line and is thus rotated by an angle $\varphi(P^*)$ from the original one. The origin of this frame is defined at $(x(s=W/2), y(s=W/2))$ the junction point, which we note as $(x_j, y_j)$.

\noindent
Expressing in ($\mathbf{e_x^*},\mathbf{e_y^*}$) the coordinates of a point $M(x_M, y_M)$ originally expressed in ($\mathbf{e_x},\mathbf{e_y}$), we obtain:

\begin{equation}
    \mathbf{OM} = ((x_M - x_j)\cos\varphi + (y_M - y_j)\sin\varphi)\mathbf{e_x^*} + (-(x_M - x_j)\sin\varphi + (y_M - y_j)\cos\varphi)\mathbf{e_y^*}
\end{equation}

The line defined by the axis $\mathbf{e_y^*}$ is an axis of symmetry of the two cells. There is an overlap between the two shapes, and thus contact, if any point $M$ on the left tube is such that the component of $\mathbf{OM}$ held by $\mathbf{e_x^*}$ is positive. This condition gives the threshold to determine the pressure of first contact. $P_{\mathrm{contact}}$ is the minimum pressure for which there is one point $M$ (other than the junction itself) such that:

\begin{equation}
    (x_M - x_j)\cos\varphi + (y_M - y_j)\sin\varphi \geq 0.
\label{threshold_eq_sym_contact}
\end{equation}

\begin{figure}[h!]
    \centering
    \includegraphics[width = 0.9\textwidth]{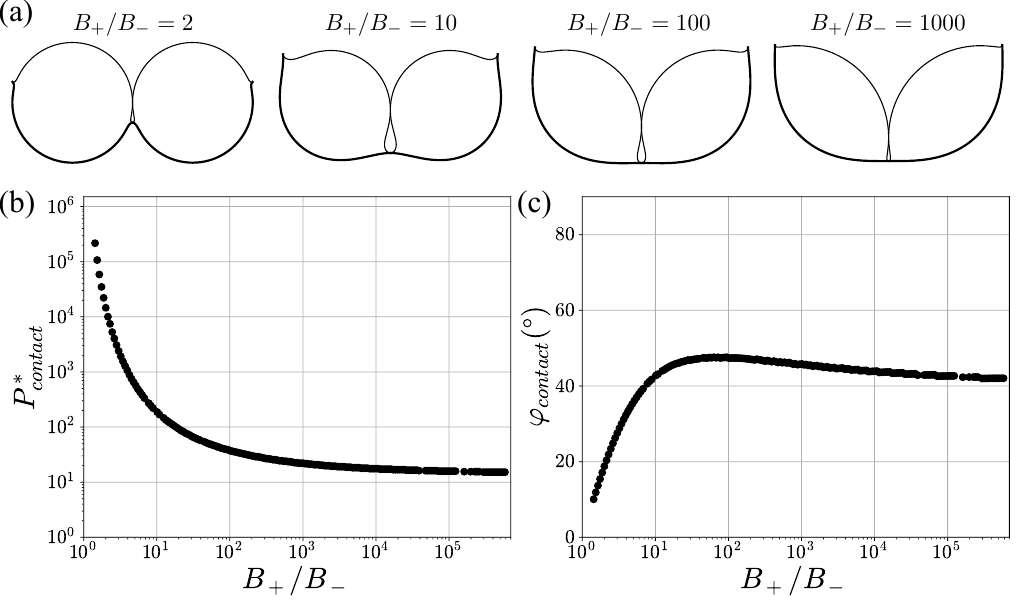}
    \caption{(a) Profiles of the cross-section at the lowest pressure at which contact appears for increasing ratios of asymmetry. (b) Evolution of $P^*_{\mathrm{contact}}$ with $B_+/B_-$. (c) Evolution of $\varphi_{\mathrm{contact}}$ with $B_+/B_-$.}
\label{tubes:contact_sections}
\end{figure}

Each term of this equation depends on the pressure, and it provides a threshold to determine whether contact has occurred or not. We solve numerically the systems defined by Eqs.~\ref{tube_kirchhoff_plus_adim} and Eqs.~\ref{tube_kirchhoff_moins_adim} by slowly increasing the pressure, checking the threshold at each step. We stop the solution and record the pressure as $P_{\mathrm{contact}}$ when the threshold condition Eq.~\ref{threshold_eq_sym_contact} is first met. We show a few cross-sections in Fig.~\ref{tubes:contact_sections}, as well as the evolution of $P_{\mathrm{contact}}$ and $\varphi_{\mathrm{contact}}$ with the stiffness asymmetry.

\clearpage

\paragraph{Accounting for the width of the seam line.} The previous discussion considers the junction of the tubes as point-like. Before contact, the seam line remains straight as no torque is applied. We can account for the width of the seam line $e_\mathrm{seam}$ in the threshold to determine the contact. Eq.~\ref{threshold_eq_sym_contact} becomes:

\begin{equation}
        (x_M - x_j)\cos\varphi + (y_M - y_j)\sin\varphi \geq \frac{e_\mathrm{seam}}{2}.
\end{equation}

\begin{figure}[h!]
    \centering
    \includegraphics[width = 1\textwidth]{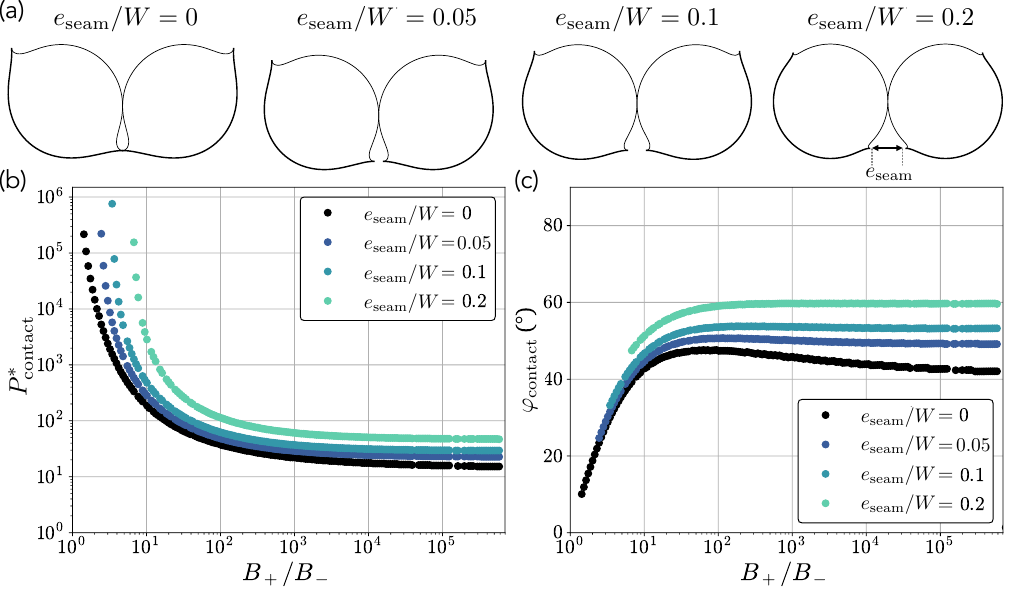}
    \caption{(a) Profiles of the cross-section at lowest pressure at $P^*=P^*_{\mathrm{contact}}$ for increasing relative seam line width $e_{seam}/W$ for $B_+/B_-=10$. (b) Pressure at which contact occurs for two identical tubes with an infinitely thin seam line. (c) Corresponding junction angle at first contact.}
\label{tubes:contact_eseam_graphs}
\end{figure}

We show in Fig.~\ref{tubes:contact_eseam_graphs} (a) a few cross-sections at the time of the first contact for $B_+/B_-=10$ and increasing values of the width of the seam line $e_{\mathrm{seam}}$. At a given ratio of asymmetry, the contact pressure increases with the width of the seam. Fig.~\ref{tubes:contact_eseam_graphs} (b-c) shows the evolution of $P_{\mathrm{contact}}^*$ and $\varphi_{\mathrm{contact}}$ for increasing ratios of stiffness asymmetry. The threshold for contact $P_\mathrm{contact}$ increases with the width of the seam line, here made dimensionless using the width of the tube. We do not see any significant variation of the threshold in the experimental data reported in Fig. 3 of the main paper as experiments in the contact regime were performed only on tubes of widths between 2 and \SI{5}{\centi\meter}, i.e. $0.002<e_{\mathrm{seam}}/W < 0.05$. The width of the seam line $e_{\mathrm{seam}}$, difficult to precisely assess, is of the order of one millimeter. In practice, the seam line would have an effect on the contact only in the case of ratios $e_{\mathrm{seam}}/W > 0.1$.

\begin{figure}[h!]
    \centering
    \includegraphics[width = 1\textwidth]{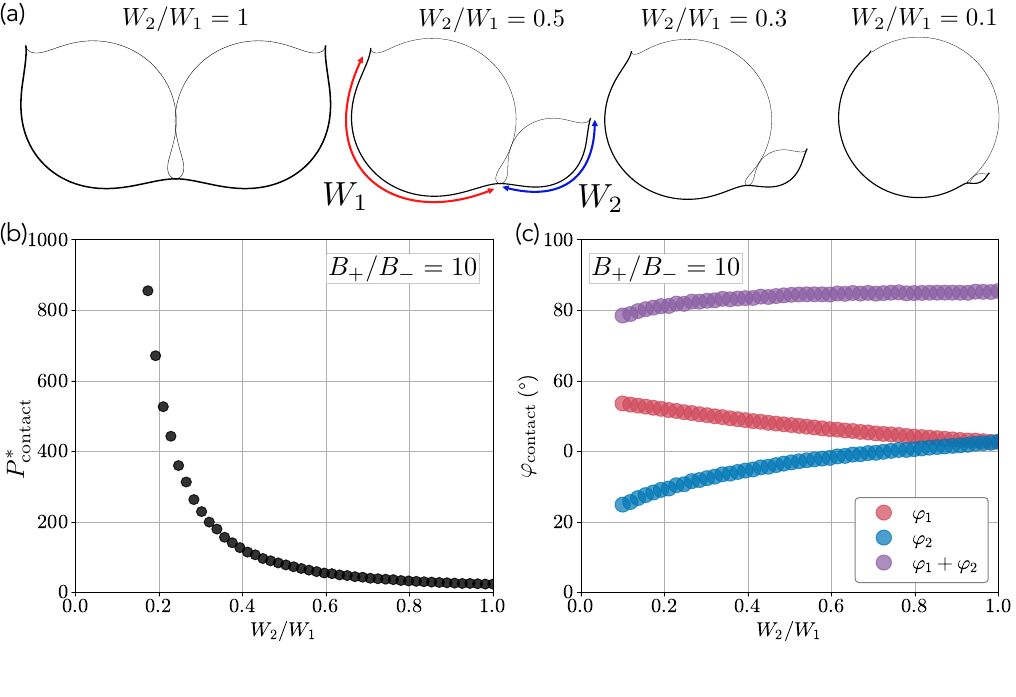}
    \caption{(a) Profiles of the cross-section at lowest pressure at which contact appears for increasing asymmetry of widths, for $B_+/B_-=10$.  (b) Pressure at which contact occurs for two identical tubes with an infinitely thin seam line. (c) Corresponding junction angle at first contact.}
\label{tubes:contact_width_asym}
\end{figure}

\paragraph{Neighboring tubes differing in width.} We can extend our analysis to neighboring tubes which differ in widths. In this case, we cannot use as easily the axis of symmetry between the tubes. We must then compute the distance between each point of each tube at a given pressure, and determine whether they are in contact or not. This approach provides predictions as showcased in Fig~\ref{tubes:contact_width_asym} (a) through a few examples. We report in Fig~\ref{tubes:contact_width_asym} (b) the pressure at first contact for a given ratio of $B_+/B_-$, with the pressure non-dimensionalized using the wider tube. The larger the asymmetry, the more the large tube must be inflated with regard to its own $P^*$. We similarly report in Fig~\ref{tubes:contact_width_asym} (c) the evolution of the two junction angles. It is notable that their sum appears to be almost constant.

\vspace{3mm}

\noindent {\large {\textsc{(2). FEM model}}}
\vspace{3mm}

To obtain a quantitative prediction of the re-opening which follows contact, we perform finite element simulations through the proprietary software ComSol Multiphysics 5.6. We note that we could have adapted our Kirchhoff beam model to account for contact. The contact area can be assumed to be straight as it is subject to equal pressure on both sides, but when solving the system we do not know the position or length and area of this zone. This could be done using a double shooting method, and there are numerous works in the literature which discuss methods of doing so \cite{domokos1997constrained,roman2002postbuckling,chen2014deformation,batista2017large, virgin2018tailored, chen2020contact}. Since there is an axis of symmetry between the tubes, we know the orientation of the contact zone from the junction angle. It is related to the constrained \textit{elastica}, only with the constraining plane rotating based on the inflating pressure. The problem appears tractable numerically with our simple tools.
Nonetheless, we did not deem the algorithmic complexity of this effort worth it for our practical purposes. We thus use the finite element model (FEM) ComSol in this section.

Our approach was inspired by the one of Andrade-Silva and Marthelot~\cite{andrade2023fabric}. We modeled two sheets bound together using the shell interface included in the Structural Mechanics Module. The mechanical model accounted for linear elastic Hookean materials, with geometric nonlinearities.

To model the presence of seams in the physical system, the geometry is represented by two independent surfaces that are connected along three prescribed lines. These seam lines include two along the lateral edges and one along the central axis of the structure. The central seam is additionally constrained to reproduce the experimentally relevant boundary condition of a fixed interface. The use of independent surfaces joined along these lines allows for a more flexible and realistic description of the mechanics of the plates, including realistic application of pressure over each face and the discontinuities in each seam line.
Contact between surfaces is modeled using a frictionless formulation, thereby neglecting tangential forces and focusing solely on normal interaction. This assumption is justified by the dominant role of normal contact forces in the configuration under consideration and simplifies the numerical treatment of the problem. We performed preliminary tests with various friction coefficients and found that the global behavior did not change; therefore, we neglected this effect in the numerical experiments. The contact algorithm is implemented to prioritize computational efficiency: instead of performing a global search for all possible contact interactions, it is restricted to regions where contact is expected.

Specifically, contact detection is limited to surfaces characterized by lower bending stiffness. Since these regions are more prone to deformation, they are identified a priori as the most likely locations for contact initiation. By confining the contact search to these subsets of the model, the number of candidate interactions is significantly reduced, thereby substantially decreasing computational cost without compromising accuracy. This targeted strategy enables efficient simulation of the system while preserving the essential mechanical features that govern contact behavior. This same approach was also used to numerically model the system stiffness by applying an external force to the side seams at different inflation pressures. We have checked that the scaling of geometrical parameters in $PW^3$ holds in these simulations.

The behavior is qualitatively similar to our experimental observations: the angle between the tubes starts to decrease after contact, with a small area of contact appearing and migrating toward the junction downward as the pressure increases. It is notable that the contact area remains very small with regard to the width of the tube and to the junction area. In Fig. 3 of the main paper, we plot the prediction of the FEM model along experimental data, with good agreement. In Fig.~\ref{tubes:contact_fin}, we report the evolution of the junction angle for different stiffness asymmetries and we do observe a slow decrease of $\varphi$ after contact is reached. In this plot, it appears that regardless of the stiffness asymmetry, the data collapse onto a single curve at sufficiently high pressures. We also plot, as an example, the results for a system with low stiffness asymmetry for which contact does not occur at the dimensionless pressures explored. We have checked that, before contact, the FEM prediction of the junction angle agrees with the Kirchhoff rods model.

\vspace{3mm}
\noindent {\large {\textsc{(3). Geometric model}}}
\vspace{3mm}

\begin{figure}[h!]
    \centering
    \includegraphics[width = 0.6\textwidth]{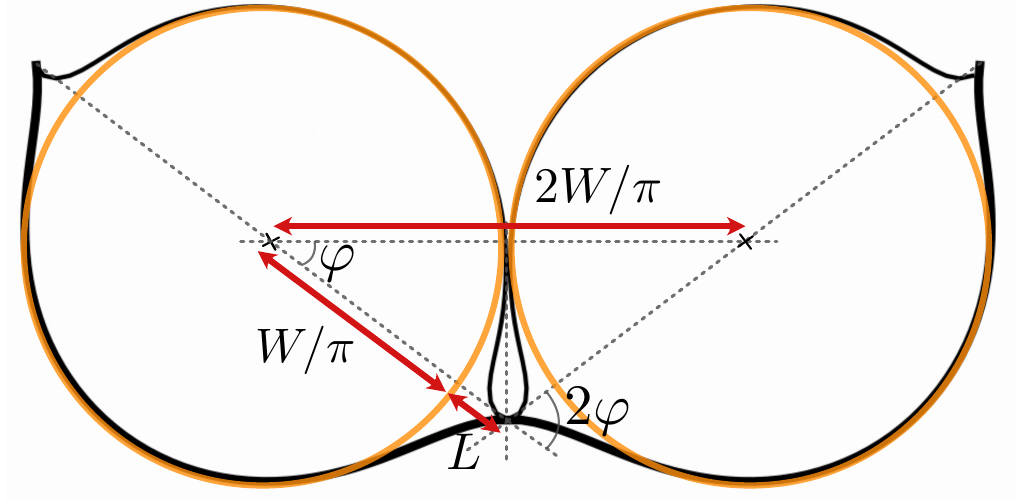}
    \caption{Geometrical construction to determine the rotation at large pressures.}
\label{tubes:contact_scaling}
\end{figure}

\begin{figure}[h!]
    \centering
    \includegraphics[width = 0.8\textwidth]{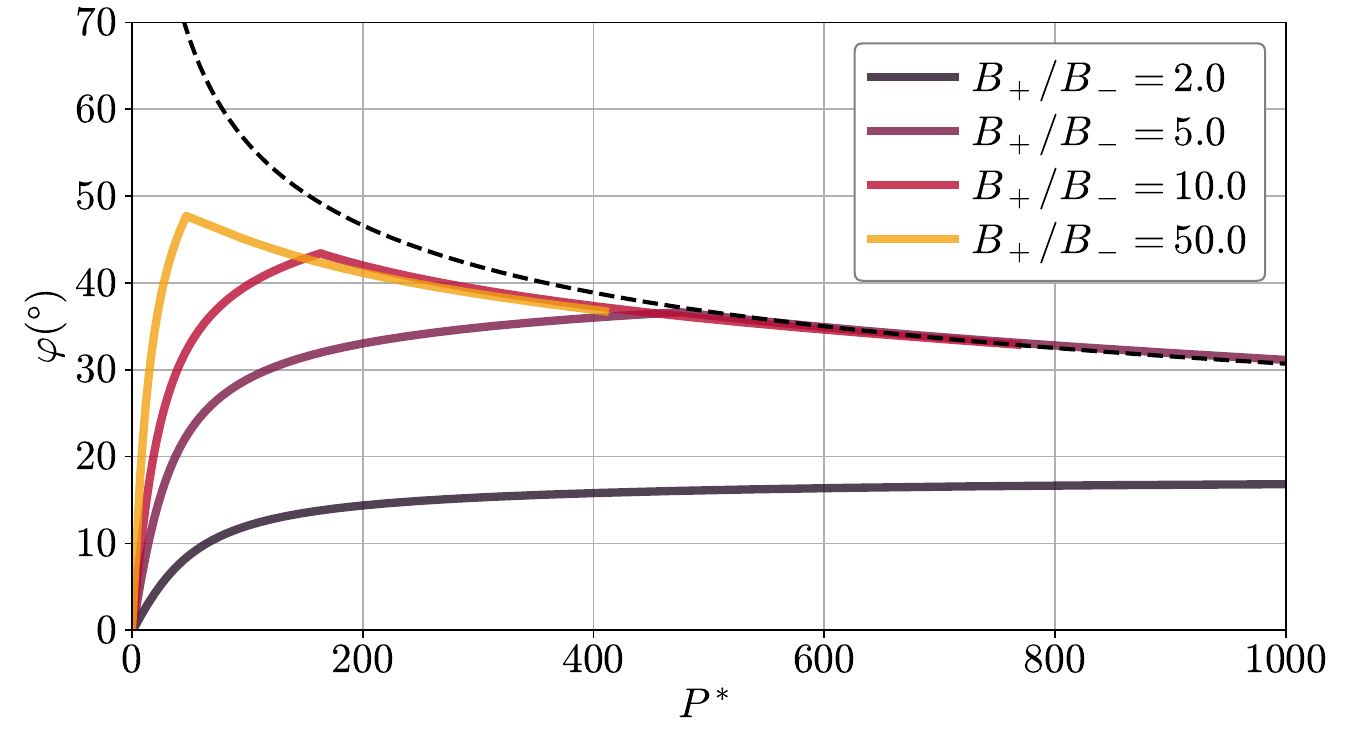}
    \caption{Evolution of the junction angle as a function of the dimensionless pressure obtained through FEM simulations for various degrees of asymmetry. The dashed black line corresponds to Eq.~\ref{contact:eq_scaling}.}
\label{tubes:contact_fin}
\end{figure}

We propose here a simplified scaling model which accounts for the post-contact evolution of $\varphi$ at large pressures. In Fig.~\ref{tubes:contact_scaling}, we sketch two tubes in contact. For sufficiently large pressures, we assume that the cross-sections are mostly circular (orange circles), of diameter $2W/\pi$. Since the contact area appears small in FEM simulations, we assume that the two circles are intersecting over a single point. The junction angle is related to the length between the center of the circles to the junction -- which we assume infinitely thin.

\noindent
Expressing the cosine of the angle shown on the left of the sketch, we get:

\begin{equation}
    \cos \varphi = \frac{W/\pi}{W/\pi + L},
\end{equation}

\noindent where $L$ is the distance from the junction point to the circle fitting the cross-section.

The size of the areas that deviate from the circular cross-section scales as $\sqrt{B_\pm/N}$ with $N$ the tension in the membrane. In the regime of high pressures, the membrane tension is equal to $PW/\pi$. Let us assume that the distance $L$ is equal to this scaling with respect to the stiffer membrane. We thus get an expression for the junction angle after contact at high pressures:\

\begin{equation}
    \cos \varphi = \frac{W/\pi}{W/\pi + \sqrt{B_+\pi/PW}}.
\end{equation}

\noindent which we write as a function of the dimensionless pressure:

\begin{equation}
    \cos \varphi = \frac{1}{1 + \sqrt{\pi^3/P^*}}.
\label{contact:eq_scaling}
\end{equation}

We plot the result from FEM simulations along Eq.~\ref{contact:eq_scaling} in Fig.~\ref{tubes:contact_fin} -- we observe a convergence of the behavior at large pressures regardless of the ratio of asymmetry. We find a great agreement at large pressures ($P^* > 500$). It is surprising that no fitting parameter is needed for the collapse, as we used a scaling argument. A physical interpretation of the behavior at high pressures is that the lever arm associated to the counter torque applied by the contact zone scales as the square root of the pressure, with the force scaling linearly with the pressure. This leads to a relatively slow decrease which depends on the square root of the pressure.

\clearpage

%%%%%%%%%%%%%%%%%%%%%%%%%%%% UNCORRELATED EVENTS %%%%%%%%%%%%%%%%%%%%%%%%%%%% 

\noindent {\Large {\textsc{F. Bending rigidity of tubes}}}

\begin{figure}[h!]
    \centering
    \includegraphics[width = 1\textwidth]{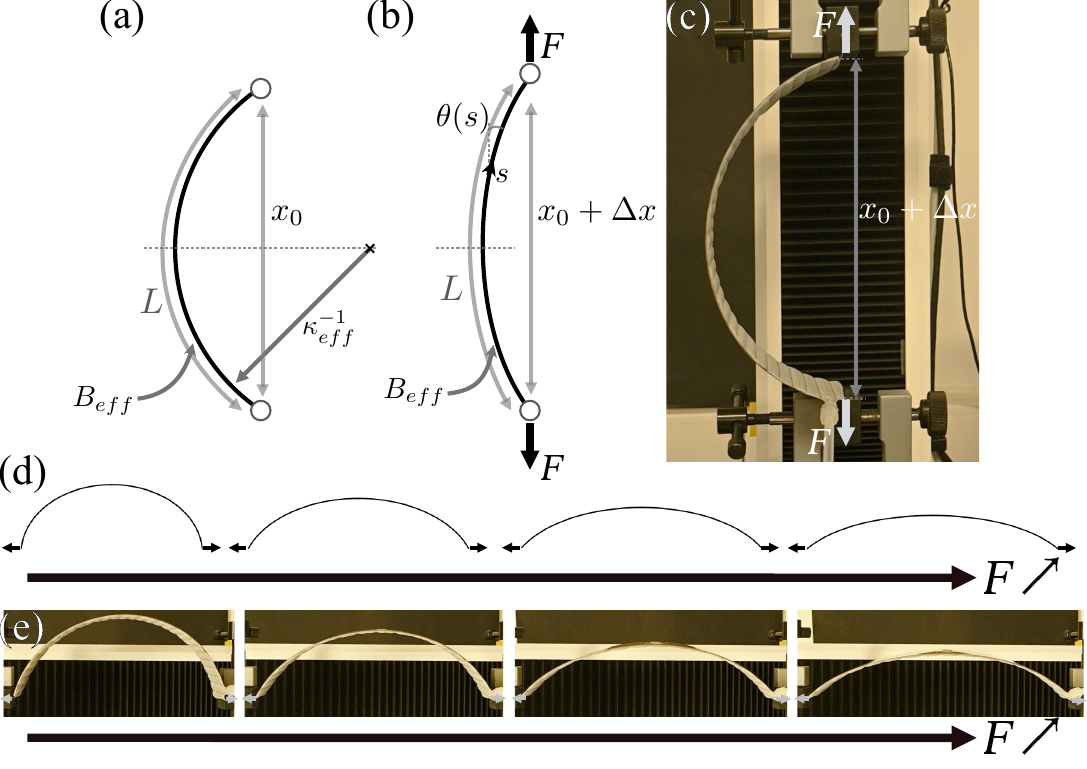}
    \caption{(a-b) Idealized network in an inflated state as a rod with no force applied (a) and during the pulling (b). (c) Corresponding experimental picture of the network (here made of $n=30$ tubes of width $W=$ \SI{1}{\centi\meter} with a stiffness asymmetry $B_+/B_-=800$). (d-e) Visualization of numerical solution of the curved \textit{elastica} with increasing forces (d) and corresponding experimental images (e).}
\label{tubes:stiffness_manip1_illu}
\end{figure}

\paragraph*{Low pressure regime.} Considering a network of tubes inflated at a low pressure, one can conceptualize the inflatable itself as a rod with a constant intrinsic curvature $\kappa_{0}$ -- a function of $P^*$ and $B_+/B_-$. With $L$ the length of the network, the dimensionless number $\kappa_0L$ fully describes the shape as it corresponds to a portion of a circle (e.g. for $\kappa_{0}L = \pi$ the rod forms a half-circle). Contact between the two end points occurs at $\kappa_0 L = 2\pi$. A sketch of a rod with no external loading applied is shown in Fig.~\ref{tubes:stiffness_manip1_illu} (a).

We study the deformation of the rod in a pinned-pinned configuration as it is pulled by two coaxial forces of norm $F$ along the direction joining its two ends while the inner pressure is maintained, as shown in Fig.~\ref{tubes:stiffness_manip1_illu} (b) and (c). This, in practice, uncurls the inflatable. Since we apply a force rather than a pure torque, the curvature of the rod is no longer homogeneous for $F > 0$. In the assumption of inextensibility, the rod fully straightens for $F \to +\infty$. We note the distance between the two end points in the rest state as $x_0$ and its increase as $\Delta x$. The experiments were made using a traction machine in which the value of $x_0 + \Delta x$ is controlled and the applied force $F$ is measured. Sequential pictures for increasing forces are shown in Fig.~\ref{tubes:stiffness_manip1_illu} (d) and (e).

\begin{figure}[h!]
    \centering
    \includegraphics[width = 1\textwidth]{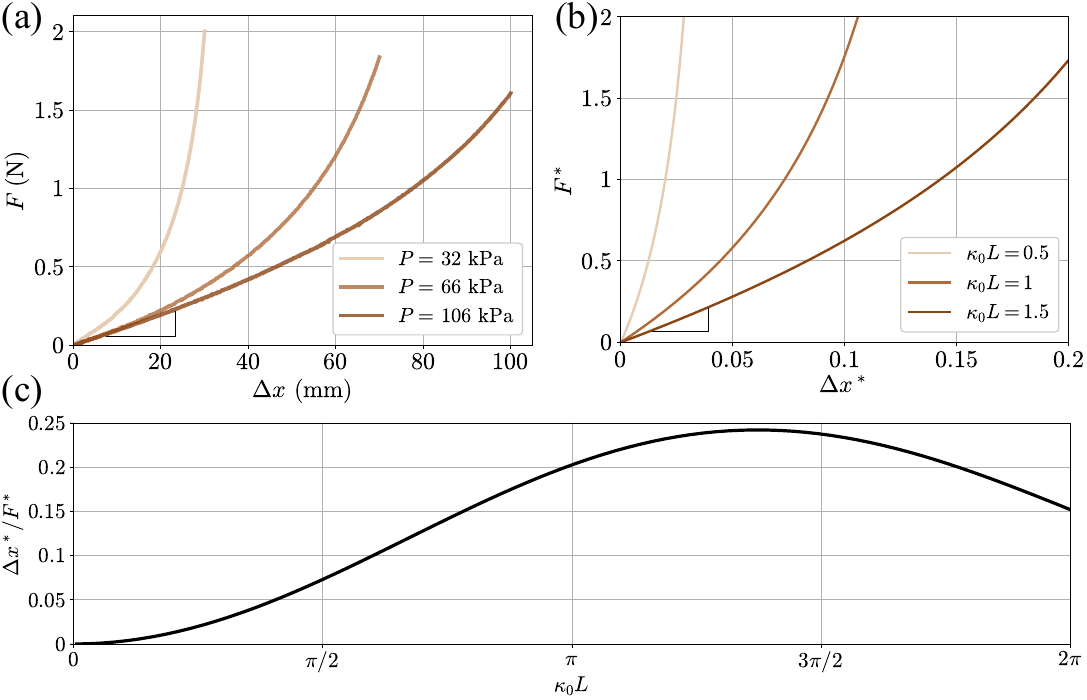}
    \caption{(a) Experimental force-displacement curves obtained for a given network at three different inflating pressures (here made of $n=30$ tubes of width $W=$ \SI{1}{\centi\meter} with a stiffness asymmetry $B_+/B_-=800$). (b) Force-displacement curves obtained by numerical resolution of the curved rod pulled by two forces at different dimensionless shape parameter $\kappa_0L$. (c) Slope of the force-displacement linear response as a function of the shape parameter $\kappa_0L$ as predicted by our numerical resolutions such as the ones shown in (b).}
\label{tubes:stiffness_manip1_expe}
\end{figure}

In practice, we first pressurized the network of tubes and measured its curvature away from the traction machine. The inflatable is fitted on both ends with floppy inextensible fins -- similar to the ones used to measure the junction angle but less stiff. We then clamped the floppy fins %of the inflatable 
into the traction machine. The inextensible fins transmit forces but no torque to the end points of the network. Since the edge angles are not set, we can consider that the network is in a pinned-pinned situation. We pull it with two opposite forces in a quasi-static manner, and record the evolution of the distance $\Delta x$ with the increasing force $F$. Examples of force-displacement curves are shown in Fig~\ref{tubes:stiffness_manip1_expe} (a). It is difficult to determine precisely the actual origin of the curve due to the weight of the inflatable: even when no force is applied they are deformed by gravity. For a given sample, we perform the experiments several times while varying the pressure value between experiments. We do several sets of experiments for networks of different widths and different stiffness asymmetries. Note that this method is limited to objects that have an intrinsic curvature -- we cannot use it to probe the symmetric case ($B_+/B_- = 1$), nor can it be used for networks with very large curvatures due to self-contact. We are thus limited to the study of the behavior at low dimensionless pressures.

Our objective is to rationalize the small deformation regime of the behavior. However, we cannot directly infer the effective bending stiffness of the network $B_{\mathrm{eff}}$ from the linear slope of these curves. Indeed, the relation between $\Delta x$ and $F$ for a given bending stiffness $B_0$ depends on the shape of the rod at rest, i.e. on the dimensionless shape parameter $\kappa_0L$. We must then determine the theoretical relationship between the force and the displacement at a given bending stiffness.

As the only external forces are applied at the extremities of the network, the relevant minimal model is the \textit{elastica} with an intrinsic curvature. The effects of an intrinsic curvature are seen only in the boundary conditions of the problem. We write the associated \textit{elastica} problem, solved in half of the system by symmetry with $s$ varying between $0$ (midway point) and $L/2$ (free end):

\begin{equation}
    B_{\mathrm{eff}}\dfrac{\dd^2 \theta}{\dd s^2} - F\sin \theta = 0,
\end{equation}

\noindent with the two boundary conditions:

\begin{equation}
    \begin{cases}
        \theta(0) = 0, \\[8pt]
        \dfrac{\dd \theta}{\dd s}(L/2) = \kappa_0.
    \end{cases}
\end{equation}

\noindent
Using $L$ as a length scale, the dimensionless version of the equation writes:

\begin{equation}
    \dfrac{\dd^2 \theta} {\dd s^{*2}} - \frac{FL^2}{B_{eff}}\sin \theta = 0,
\label{tube:elastica_stiff}
\end{equation}

\noindent with the two boundary equations:

\begin{equation}
    \begin{cases}
        \theta(0) = 0, \\[8pt]
        \dfrac{\dd \theta}{\dd s^*}(1/2) = L\kappa_{0}.
    \end{cases}
\end{equation}

In the corresponding experimental system, only the parameters $B_{\mathrm{eff}}$ and $\kappa_{0}$ vary with the pressure in the tubes. We measure $\kappa_0(P^*)$ independently of the traction experiment to determine $B_{\mathrm{eff}}$. In our model, we determine the variation in the distance between the two end points $\Delta x^*$ (which is normalized by $L$) at a given applied force $F > 0$ as:

\begin{equation}
    \Delta x^*(F^*) = \int_{0}^{1/2} \cos \theta \dd s^* -  \int_{0}^{1/2} \cos (\kappa_0 s^*) \dd s^*.
\end{equation}

The \textit{elastica} Eq.~\ref{tube:elastica_stiff} can be solved to obtain the relation between $\Delta x^*$ and $F^* \equiv FL^2/B_{\mathrm{eff}}$. Out of simplicity, we solve the system numerically using as a boundary value problem. We show in Fig.~\ref{tubes:stiffness_manip1_expe} (b) examples of the evolution of the distance $\Delta x^*$ as a function of the pulling force $F^*$ -- the theoretical equivalent to our experimental force-displacement curves. Since we restrict our study to the linear behavior at small deformations, we extract the linear coefficient from the curves at low forces. We plot in Fig.~\ref{tubes:stiffness_manip1_expe} (c) the evolution of the linear coefficient $\Delta x^*/F^*$ as a function of the shape factor $\kappa_0L$.

\begin{figure}[h!]
    \centering
    \includegraphics[width = 1\textwidth]{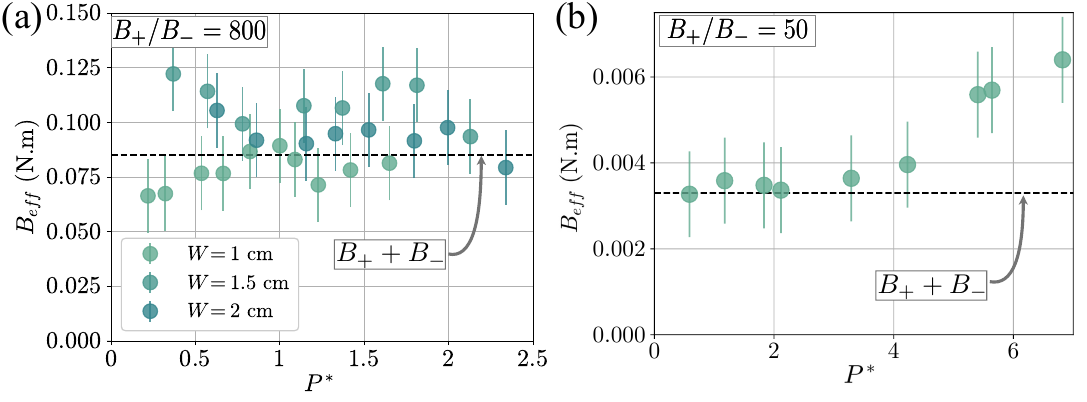}
    \caption{Evolution of the effective bending stiffness at low dimensionless pressures $P^*$ for networks of stiffness asymmetry (a) $B_+/B_-=800$ and (b) $B_+/B_-=50$. The dashed lines correspond to $B_{\mathrm{eff}}= B_+ + B_-$.}
\label{tubes:stiffness_manip1_results}
\end{figure}

It is then straightforward to obtain the effective bending stiffness of the network from our traction experiments. Experimentally, dividing the experimental measurement of the linear coefficient $\Delta x/F$ by its theoretical counterpart obtained by solving the \textit{elastica} for the same shape factor $\kappa_0L$, we obtain $L/B_{\mathrm{eff}}$. We extract in this manner the effective bending stiffness from the traction experiments, having removed the effects of geometry on the behavior. We finally plot the experimental results for $B_{\mathrm{eff}}$ in Fig.~\ref{tubes:stiffness_manip1_results} for two ratios of stiffness asymmetry at small inflating pressures. Note that the results are normalized by the length of the tubes, we thus present the effective bending stiffness per unit of tube length.

Though the method is not particularly precise, the trend is quite clear: at low dimensionless pressures, the effective bending stiffness is nearly constant. Since the pressure is low, we expect the stiffness to be a consequence of the material properties mostly. The plateau does fit fairly with the value of $B_+ + B_-$ -- note that here $B_+ + B_- \simeq B_+$. In the case of $B_+/B_- = 50$, we see a transition from this saturation plateau to a slow increase around $P^* = 5$. We nonetheless need to use another characterization method to be able to probe inflated networks with larger curvatures, as we are strongly limited by the limitation of the shape parameter $\kappa_{0}L < 2\pi$.

Our method provides a measurement of the bending stiffness in the case of low intrinsic curvatures ($\kappa_0L < 2\pi$). The result is thus particularly relevant to networks of tubes with a large stiffness asymmetry ($B_+/B_- > 100$). Another method, working on systems of two tubes only, is discussed in the main paper to characterize the rigidity of asymmetric inflatables at larger dimensionless pressures.

\begin{figure}[h!]
    \centering
    \includegraphics[width = 0.8\textwidth]{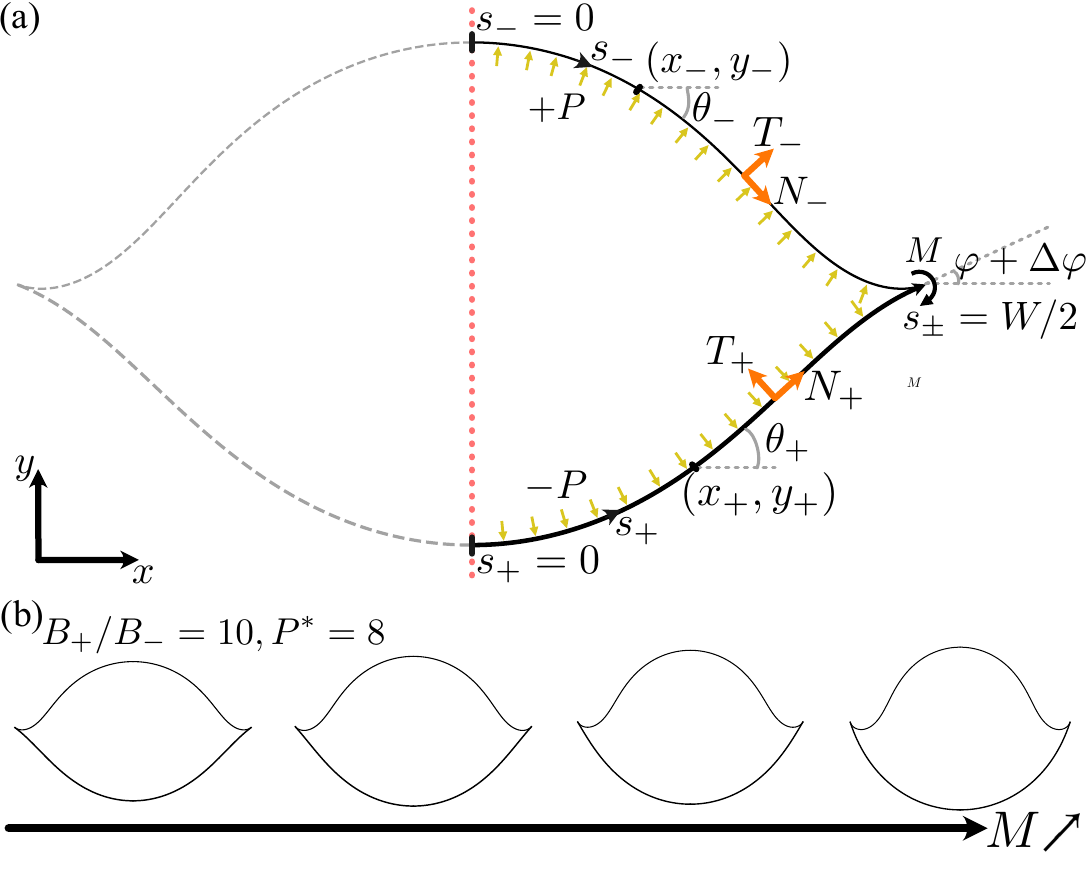}
    \caption{(a) Sketch of the coupled Kirchhoff rods model with an applied point torque at the junction. (b) Cross-sections obtained from numerical resolutions for $B_+/B_- = 10$, $P^* = 10$ as the norm of the applied torque is increased. Note that the applied torque is negative here.}
\label{tubes:kirchhoff_moment}
\end{figure}

\clearpage
\paragraph{Modeling with a single cross-section.} We propose to adapt the coupled Kirchhoff rods model used to model the geometry of the cross-section to the characterization of its bending stiffness. Fig.~\ref{tubes:kirchhoff_moment} (a) shows a sketch of the modeled system.

\begin{figure}[h!]
    \centering
    \includegraphics[width = 0.85\textwidth]{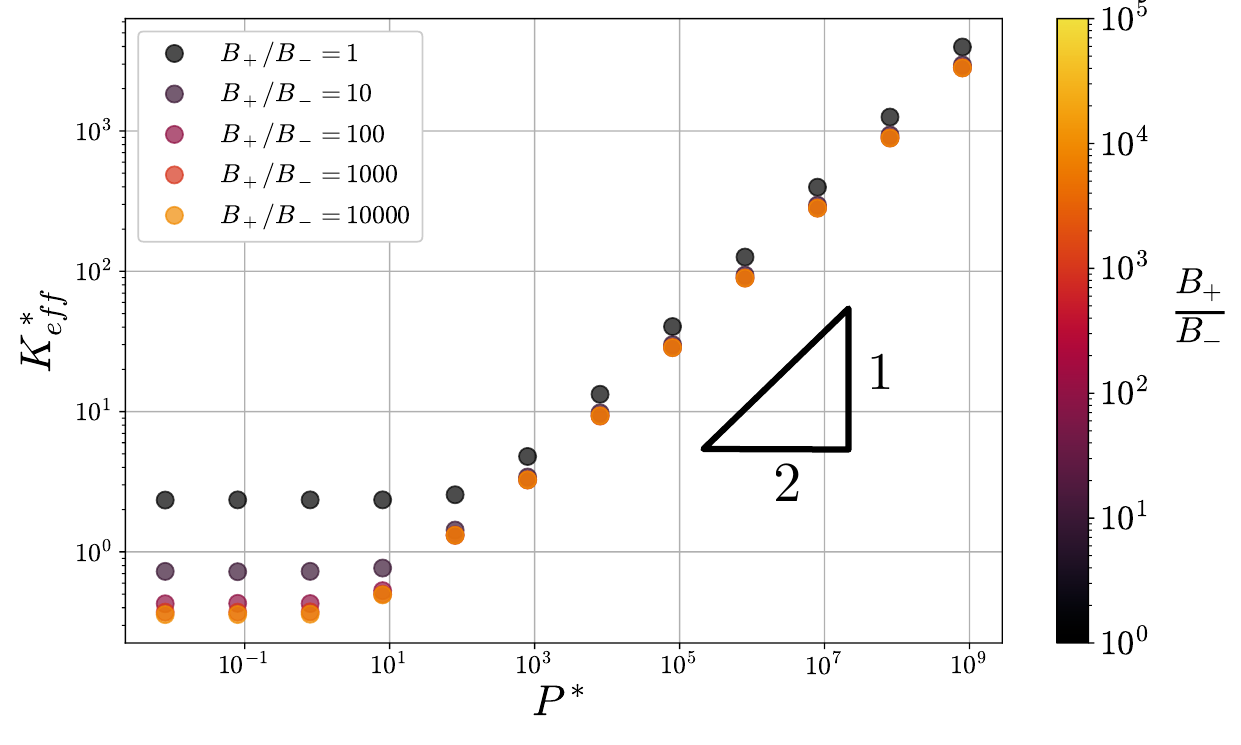}
    \caption{Evolution of the rotational stiffness $K^*_{\mathrm{eff}}$ as a function of the dimensionless pressure $P^*$ as predicted by the Kirchhoff rod model.}
\label{tubes:kirchhoff_moment_result}
\end{figure}

We consider the case of a point torque applied at the junction point. As such, the Kirchhoff rod equations (Eqs.~\ref{tube_kirchhoff_plus_adim} and~\ref{tube_kirchhoff_moins_adim}) are unchanged, and we still model half of the system by symmetry, between the midpoint at $s^*=0$ and the junction point at $s^*=W/2$. For the boundary conditions (Eqs.~\ref{tube:boundary_condition_midpoint} and~\ref{tube:boundary_condition_junction}), only the condition for the mechanical coupling of the torques is changed and becomes:

\begin{equation}
    \kappa^*_+(s_+^*=1/2) + \frac{B_-}{B_+}\kappa^*_-(s^*_-=1/2) = M^*,
\end{equation}

\noindent with $M^*$ the point torque normalized by $B_+W$. Fig.~\ref{tubes:kirchhoff_moment} (b) shows an example of the response of the cross-section of a tube when a negative torque is applied at its junctions. We focus here on the linear response of the system, and thus get the linear prefactor $K_{\mathrm{eff}}^* = \delta \varphi/\delta M^*$. We plot in Fig.~\ref{tubes:kirchhoff_moment_result} their evolution for various stiffness asymmetries.

We observe two distinct regimes with a transition around $P^* = 10$ with a slight variation of the threshold based on the stiffness asymmetry. For low dimensionless pressures, the stiffness of the tube is virtually independent from $P^*$ and is a function of the stiffness of the rod. We would perhaps expect a linear dependency of the stiffness on the pressure in this regime, but since $P^*$ is very low we expect the bending resistance of the rods to dominate over the pressure. The stiffness here decreases with the ratio of $B_+/B_-$, this is due to the choice of normalization with regard to $B_+$. For larger dimensionless pressures, we observe a surprising dependency of the effective stiffness with the square root of the inflating pressure, which agrees with our experimental observations. This observation can be rationalized through the analysis of the boundary layer.

The Kirchhoff model agrees qualitatively with the experiments in reproducing the evolution of the stiffness of tubes in the regime in which there is no contact. 

\newpage

\paragraph*{Boundary layer analysis.} From the loaded cross-sections for $P^*=10$ visible in Fig.~\ref{tubes:kirchhoff_moment} (b), we observe that most of the deformation occurs near the junction for relatively large dimensionless pressures. Indeed, the deformation due to pressure dominates over the one due to the applied torque over most of the cross-section apart from the junction. The rotation between two consecutive tubes is thus only due to deformations near the junction.

\begin{figure}[h!]
    \centering
    \includegraphics[width = 0.8\textwidth]{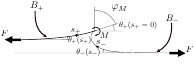}
    \caption{Junction area as two \textit{elastic\ae} with a point torque applied at the junction point.}
\label{tubes:elastica_stiffness}
\end{figure}

We can extend our analysis of the boundary layer at the junction in terms of two \textit{elastic\ae} by adding a counter torque. The system corresponds to two elastic\ae~\cite{vani2025asymmetric} with only the mechanical coupling condition changing -- similarly to the change concerning the coupled Kirchhoff bending rods. The condition is now written:

\begin{equation}
    B_+\frac{\dd \theta_+}{\dd s}(s=0) - B_-\frac{\dd \theta_-}{\dd s}(s=0) = -M.
    \label{eq:new_condit_torque_bl}
\end{equation}

The system of equations can be solved analytically. This is done in great details in~\cite{vani2025asymmetric} and is proven with well-controlled experiments on ribbons. The stiffness of the system can be linearly approximated as:

\begin{equation}
    \Delta \varphi \simeq \frac{M}{2\sqrt{N \left( B_+ + B_- \right) }}
    \label{eq:tube_boundary}
\end{equation}

The tension force for the tube scales as $P$. The boundary layer analysis captures well the trend in the large pressures regimes predicted by the Kirchhoff rod model. One particularly interesting point is that the stiffness is determined by this scaling even at relatively low dimensionless pressures ($10^2$) when the junction angle is not saturated and thus the boundary layer is not fully developed, as seen in Fig.~\ref{tubes:kirchhoff_moment_result}.

\paragraph*{FEM model.} To probe the regime in which there is contact between neighboring tubes, we again use the FEM software ComSol Multiphysics. Experiments follow the same geometry as the experiment on highly inflated tubes: two forces are applied at the edges of two connected tubes, and we extrapolate the applied torque on the junction from the associated lever arm. The simulations are done through a double continuation method, first inflating it at a given pressure and then applying an increasing amount of force.

\begin{figure}[h!]
    \centering
    \includegraphics[width = 1\textwidth]{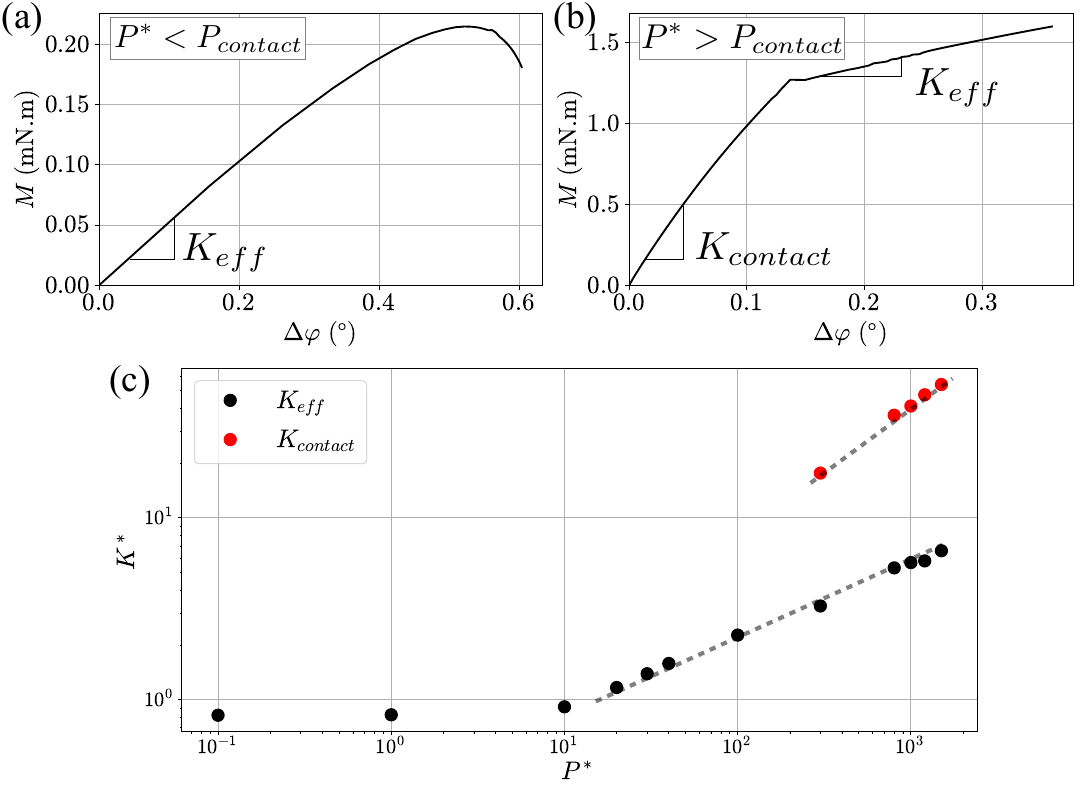}
    \caption{Results from FEM simulations. Evolution of the applied torque with the variation in junction angle for a pressure lower than the contact pressure (a) and larger (b). (c) Evolution of the dimensionless torsional stiffness as a function of the dimensionless pressure without contact (black markers) and with contact (red markers). Dashed lines indicate slopes of 0.4 (without contact), and 0.7 (with contact).}
\label{tubes:stiffness_comsol}
\end{figure}

In Fig.~\ref{tubes:stiffness_comsol} (a-b), we show experimental results for the response of the system to the loading for a pressure lower (a) and larger (b) than the contact pressure. Similarly to the experiments, there is only one regime for $P^* < P^*_{\mathrm{contact}}$ with a response at first linear. For pressures larger than $P^*_{\mathrm{contact}}$, the same distinction appears as in the experiments as well, with two regimes based on whether the contact is broken or not. For larger pressures, we observe that the behavior is not linear once contact is broken. Indeed the junction area deforms as well while the contact is being broken. This could explain the slight deviation in $K_{\mathrm{eff}}$ in prefactor pressures shown in the main paper, as the response of the boundary layer could actually be already out of the linear regime.

We extract the slopes of the linear regime to determine the FEM prediction for both $K_{\mathrm{eff}}$ and $K_{\mathrm{contact}}$. We report their evolution as a function of the dimensionless pressure in Fig.~\ref{tubes:stiffness_comsol} (c). The behavior is close to the one observed experimentally, with two power laws differing in exponent. For $K_{\mathrm{eff}}$, we find an exponent of $0.4$ -- slightly smaller than the expected square root. For $K_{\mathrm{contact}}$, we find an exponent of $0.7$ -- significantly larger than the one found for $K_{\mathrm{eff}}$. It is still sublinear and agrees with our experimental observations -- though comprehensive experimental data are missing. Note that in the contact pressure, the applied torque is still estimated at the junction -- but here is a force exchange happening away from the junction. The way we define the loading is thus biased and not much can be said from the results on $K_{\mathrm{contact}}$. The important observation is that contact highly stiffens the system, and that $K_{\mathrm{contact}}$ scales differently with the pressure.

We have not performed a quantified comparison between experiments and FEM simulations, though the order of magnitudes and trends involved are close. Note that the seam lines are not modeled in our FEM simulations, which limits the analysis for large pressures. Indeed for very large dimensionless pressures, the seam lines should become the most compliant part of the arrays of tubes.% We expect it to happen when the stiffness of the boundary layer become of the order of the stiffness of the seam line.

\clearpage

%%%%%%%%%%%%%%%%%%%%%%%%%%%% UNCORRELATED EVENTS %%%%%%%%%%%%%%%%%%%%%%%%%%%% 
\noindent {\Large {\textsc{G. Additional applications}}}

We present here the shrinking calculation and mention further applications of networks of tubes.

\paragraph{Shrinking ratio.} In Fig 5 (a) of the main paper, we present an inflatable with alternating curvature which shrinks along its main axis. The formula of its shrinking is determined geometrically from a period of $n$ bars of length $W_{\mathrm{eff}}$ connected by hinges of angle $\pi - 2\varphi$. The length $L_0$ of a period at zero pressure is $nW$. The shrinking ratio is then predicted by:

\begin{equation}
    \frac{L}{L_0} = \frac{W_{\mathrm{eff}}}{nW}\sum_{k=1}^n \cos\left( \theta_k \right),
\end{equation}

where $\theta_k = \{(2k - n)\varphi\}$. We introduce $S_n = \sum_{k=1}^n \cos\left( \theta_k \right)$, which can be expressed as the real part of a sum of complex exponentials:

\begin{equation}
    S_n = \mathrm{Re}\left( \sum_{k=1}^n e^{i(2k - n)\varphi} \right),
\end{equation}

\noindent factorizing and using the expression for a geometrical sum:

\begin{equation}
    S_n = \mathrm{Re}\left( e^{-in\varphi} e^{2i\varphi} \frac{e^{2in\varphi} - 1}{e^{2i\varphi} - 1}\right).
\end{equation}

\noindent Factorizing and using half-angle expressions:

\begin{equation}
    S_n = \mathrm{Re}\left( e^{-in\varphi} e^{e^{2i\varphi}} \frac{e^{-in\varphi} \left(e^{in\varphi} - e^{-in\varphi} \right)}{e^{i\varphi}\left( e^{i\varphi} - e^{-i\varphi}\right)} \right)
\end{equation}

\begin{equation}
    S_n = \mathrm{Re}\left( e^{i\varphi} \frac{sin\left(n\varphi \right)}{\sin \phi}\right).
\end{equation}

The shrinking ratio is thus:

\begin{equation}
    \frac{L}{L_0} = \frac{W_{\mathrm{eff}}}{nW} \cos\varphi \frac{sin\left(n\varphi \right)}{\sin \phi}.
\end{equation}

\begin{figure}[h!]
    \centering
    \includegraphics[width = 0.7\textwidth]{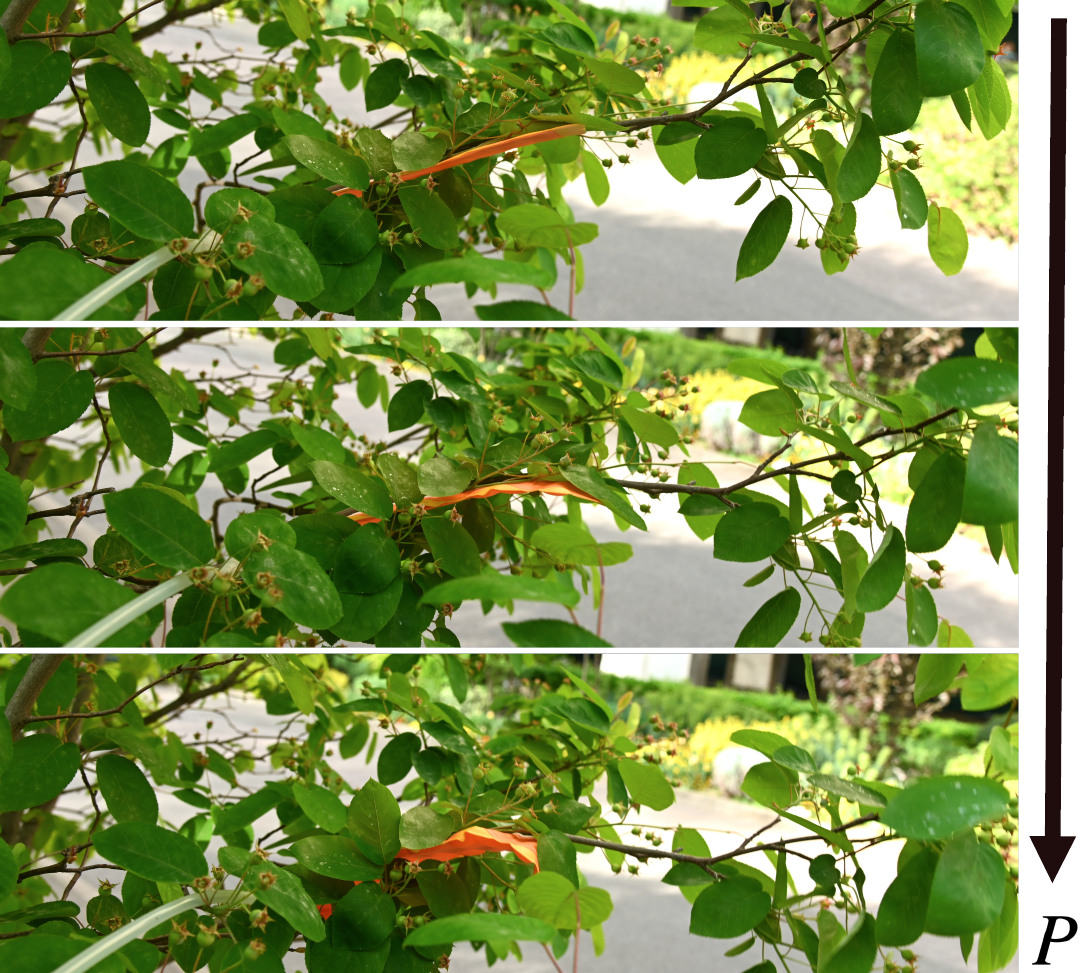}
    \caption{A symmetric network of tubes is bonded to a tree branch with double-sided tape. When the inflatable is actuated, the branch bends.}
\label{add:tree}
\end{figure}

\clearpage
\paragraph{Additional examples.} Similarly to the flower example discussed in the main paper, Fig.~\ref{add:tree} shows how the bonding of a network of tubes to a tree branch allows the remote actuation of the object.

\begin{figure}[h!]
    \centering
    \includegraphics[width = 1\textwidth]{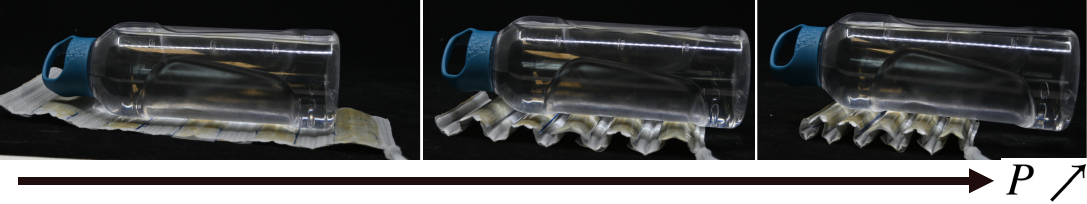}
    \caption{An inflatable with alternating curvature is used to lift a bottle filled with water, demonstrating practical load-bearing capability.}
\label{tubes:jack}
\end{figure}

In the main paper, we discuss networks of tubes with periodically alternating direction of curvature. In Fig.~\ref{tubes:jack}, we show that such structures can be used to lift objects over a small height. The rugged surface due to the out-of-plane deformation holds the bottle in place. This feature is particularly interesting in applications since the flat inflatable can be slipped below an object before inflation. Through inflation-deflation cycles, the bottle can be made to slip due to friction.

\begin{figure}[h!]
    \centering
    \includegraphics[width = 0.8\textwidth]{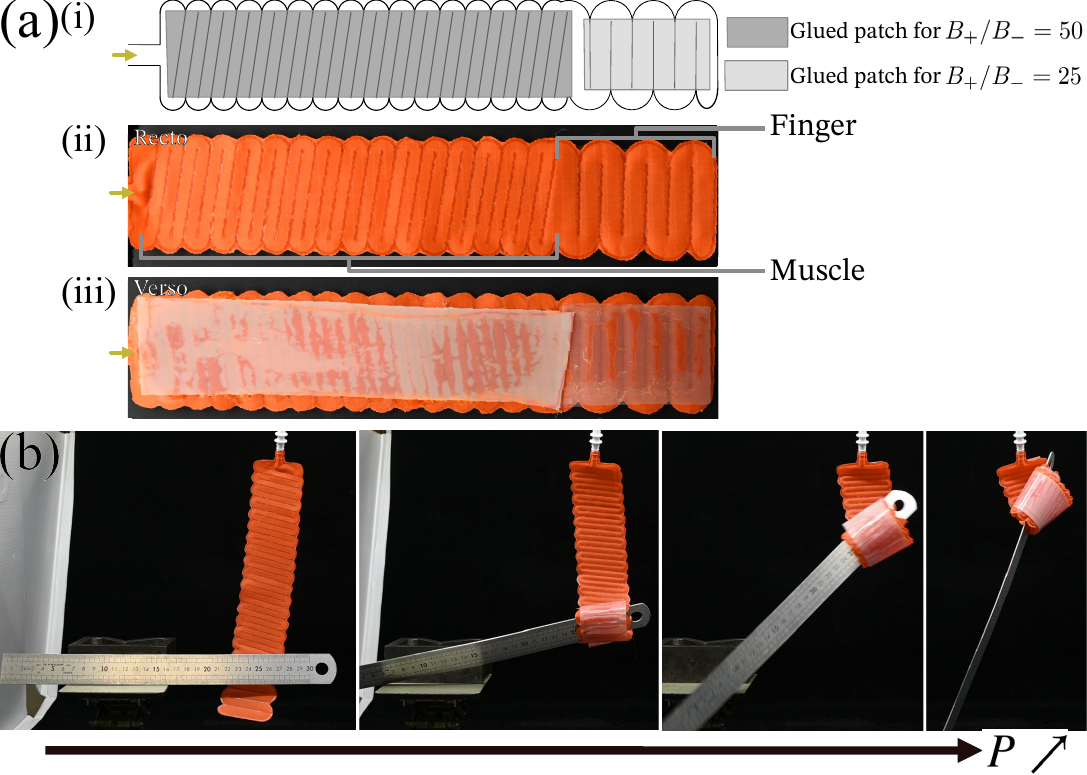}
    \caption{(a) A gripper made of two parts: a pitched muscle-like part (left) and a straight finger-like part (right). The glued patches differ in stiffness which makes the finger move at a smaller pressure than the muscle. (b) Gripper in action. Tubes are \SI{6}{\milli\meter} wide in the muscle-part and \SI{1}{\centi\meter} wide in the finger-part.}
\label{tubes:vine_gripper}
\end{figure}

We propose a refinement to the gripper shown in Fig. 4 of the main article, using a network of tubes with varying pitch angle as shown in Fig.~\ref{tubes:vine_gripper}. The rightmost part of the pattern has no pitch, and is thus easily able to grip an object -- acting as a finger. The leftmost part of the network is pitched, it thus corresponds to a muscle-like motion which will reduce the distance between the fixed point of the gripper (topmost in (b)) and the gripped ruler. Control over the stiffness of the patches gives control over $B_+$, and thus the range of working pressures for the inflatable -- the effect is enhanced by varying the widths of tubes between the two parts. The stiffening patches determine the maximal curvatures that will be reached by the different parts of the inflatables as $P_{\mathrm{contact}}^*$ varies with the ratio of asymmetry. It is generally straightforward to combine several tube patterns to program complex motions.

In Fig.~\ref{tubes:vine_gripper} (b), the gripper is designed such that it first grabs the ruler through its finger-like part, and then pulls it back with its muscle-like part. In the previous example of a gripper with a constant angle, the inflatable had to be pulled by hand to lift the object. Here the lifting is caused by the inflation of the network. A similar muscle-finger approach was implemented by coupling two distinct bubble-casted inflatables by Jones \textit{et al.}~\cite{jones2021bubble} -- but here the assembly of two inflatables is not required.

\clearpage
There are several issues with regard to industrialization of actuable, shape-morphing materials. The central challenge remains the viability of mass production. We demonstrate with a simple example that manufacturing networks of tubes at a large scale would be feasible.

\begin{figure}[h!]
    \centering
    \includegraphics[width = 1\textwidth]{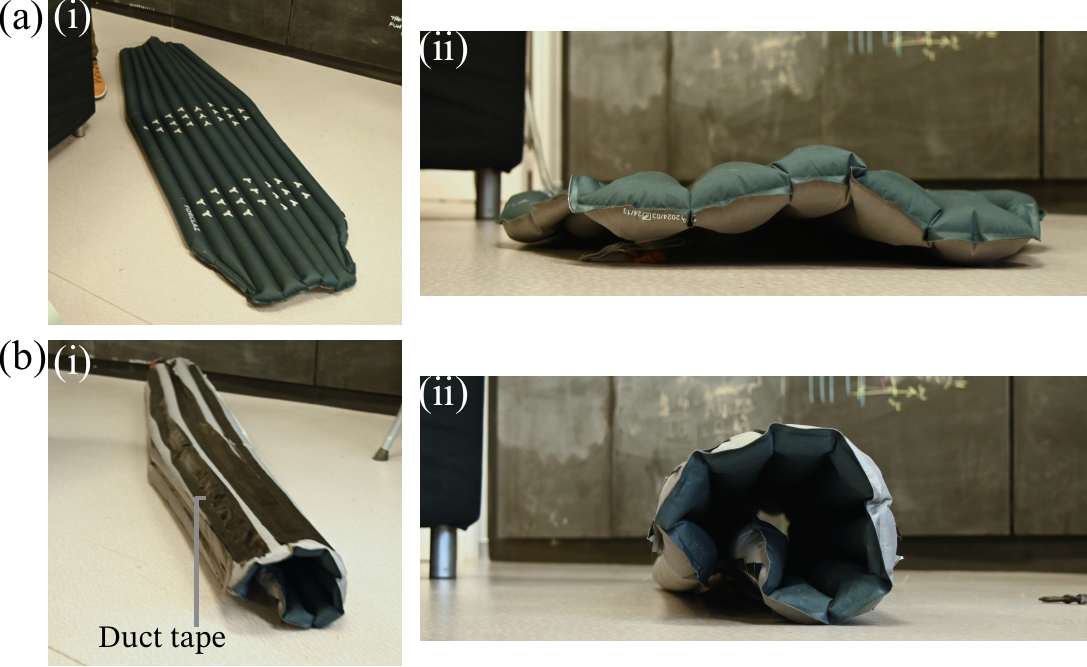}
    \caption{
    (a) Trekking inflatable mattress commercially available (Forclaz MT500). The mat is made from an network of symmetric tubes and remains mostly flat. The tubes are approximately \SI{13}{\centi\meter} wide. (b) The network is asymmetrized by application of several layers of duct tape over the soldering lines, which leads to curling upon inflation.}
\label{morphing:decath}
\end{figure}

A common observation is that the inflatables we manufacture are somewhat similar to inflatable mattresses. In Fig.~\ref{morphing:decath} (a), we show a trekking mattress commercially available from Decathlon. The mattress appears to be manufactured by heat-sealing two Nylon sheets with a press. We introduce a stiffness asymmetry through the adhesion of duct tape along the seam lines. Knowing that it is the asymmetry at the junction that governs the curvature at large dimensionless pressures, we apply four layers of duct tape over the seam lines while the mattress is deflated. We show the resulting inflated shape in Fig.~\ref{morphing:decath} (b).

We therefore believe it is possible to make networks of asymmetric tubes in large quantities -- it would also probably be easier as well to directly weld two sheets differing in stiffness.

The customized mattress also shows that the system can be scaled in size, which would be particularly important for applications in civil engineering. One issue is that the network is not particularly stiff along its direction of curvature. We have shown that this stiffness scales with the square root of the pressure, but the example shown in Fig.~\ref{morphing:decath} is still relatively soft and gravity heavily influences its shape. The issue is that the seam lines which connect the tubes are quite large, and their compliance reduces the stiffness of the tubes. This limitation is inherent to networks of tubes, but could be overcome by using layered networks of tubes~\cite{mirkin2025programming}.

\clearpage

%%%%%%%%%%%%%%%%%%%%%%%%%%%% codes  %%%%%%%%%%%%%%%%%%%%%%%%%%%% 
\noindent {\Large {\textsc{H. Codes}}}

\paragraph*{Generating soldering patterns.} We present here a Python script used to generate a Vector-Based Image in SVG format of a planar soldering pattern for a network of parallel tubes. The parameters introduced are the number of tubes, the width and length of tubes, as well as the width and length of the channel for air input. Lengths are in millimeters.

This code was written for Python version 3.11.9 using packages versions NumPy 1.24.4 and Matplotlib 3.10.0.

\begin{lstlisting}[language=Python, caption={Parametric script to generate arrays of tubes}, label={lst:python_code}]
    import numpy as np
    import matplotlib.pyplot as plt
    
    # geometrical parameters 
    length = 150 # mm
    width_channel = 15 # mm
    nbr_channel = 12 # none
    input_x_length = 7.5 # mm
    input_y_length = 6.75 # mm
    
    # creating baseline arrays for drawing lines
    nbr_point_x = 100 # nbr of points per line, any works
    x_scale = np.linspace(0, length, nbr_point_x)
    y_scale = np.zeros(nbr_point_x)
    
    # creating the figure
    f1 = plt.figure("GOOD1")
    ax = f1.add_subplot(111)
    
    def plot_twice(x, y, ax): # allow to automatically close vectors
        # allow to automatically close vectors
        vector_x = np.concatenate((x, np.flip(x)))
        vector_y = np.concatenate((y, np.flip(y)))
        ax.plot(vector_x, vector_y, 'black')
    
    # seam lines between tubes
    for row in range(nbr_channel+1):
        x = x_scale
        y = y_scale + row*width_channel
        plot_twice(x, y, ax)
        
    # arbitrary parameter of the outreach of the closing arches in mm
    parameter_arch = (width_channel-input_y_length)*1.5
    theta = np.linspace(-np.pi/2, np.pi/2, 100)
    
    # closing the last channel
    if nbr_channel%2 == 1: # odd number of channels: closed on the left
        x = -parameter_arch*np.cos(theta)
        y = width_channel/2*(1+np.sin(theta))
        plot_twice(x, y, ax)
    else: # even number of channels: closed on the right
        x = length + parameter_arch*np.cos(theta)
        y = width_channel/2*(1+np.sin(theta))
        plot_twice(x, y, ax)
        
    # arches between rows: left side
    if nbr_channel%2 == 0:
        for channel in range(int(nbr_channel/2)):
            x = -parameter_arch*np.cos(theta)
            y = width_channel*(channel*2)
                + width_channel*(1+np.sin(theta))
            plot_twice(x, y, ax)
    if nbr_channel%2 == 1:
        for channel in range(int(nbr_channel/2)):
            x = -parameter_arch*np.cos(theta)
            y = width_channel*(1+channel*2)
                + width_channel*(1+np.sin(theta))
            plot_twice(x, y, ax)
    
    # arches between rows: right side
    if nbr_channel%2 == 0:
        for channel in range(int(nbr_channel/2 - 1)):
            x = length + parameter_arch*np.cos(theta)
            y = width_channel*(1+channel*2)
                + width_channel*(1+np.sin(theta))
            plot_twice(x, y, ax)
    
    if nbr_channel%2 == 1:
        for channel in range(int(nbr_channel/2)):
            x = length + parameter_arch*np.cos(theta)
            y = width_channel*(channel*2)
                + width_channel*(1+np.sin(theta))
            plot_twice(x, y, ax)
            
    # drawing air input
    x_input_top = np.linspace(length, length + input_x_length)
    y_input_top = np.ones(np.shape(x_input_top))*(width_channel*(nbr_channel))
    plot_twice(x_input_top, y_input_top, ax)
    plot_twice(x_input_top, y_input_top - input_y_length, ax)
    # closing the gap between air input and second row
    x = length + parameter_arch/2*np.cos(theta)
    y = width_channel*(nbr_channel-1)
        + (width_channel-input_y_length)/2
        + (width_channel-input_y_length)/2*np.sin(theta)
    plot_twice(x, y, ax)
    
    # correction to the scale, removing white space
    # this part is taken from the algorithm of Panetta et al. (2021)
    def mm_to_inch(mm):
        return np.array(mm) / 25.4
    
    plt.axis('off')
    w = -np.subtract(*plt.gca().get_xbound())
    h = -np.subtract(*plt.gca().get_ybound())
    plt.subplots_adjust(0, 0, 1, 1, wspace=0.0, hspace=0.0)
    plt.gcf().set_size_inches(*mm_to_inch([w, h]))
    
    # dpi=96 is arbitrary and necessary to have the proper scale
    plt.savefig(f'tubes_serie_length={length}_width={width_channel}_nbr-channel={nbr_channel}.svg', dpi=96)
\end{lstlisting}

\begin{figure}[h]
    \centering
    \makebox[\textwidth]{\includegraphics[width=198mm]{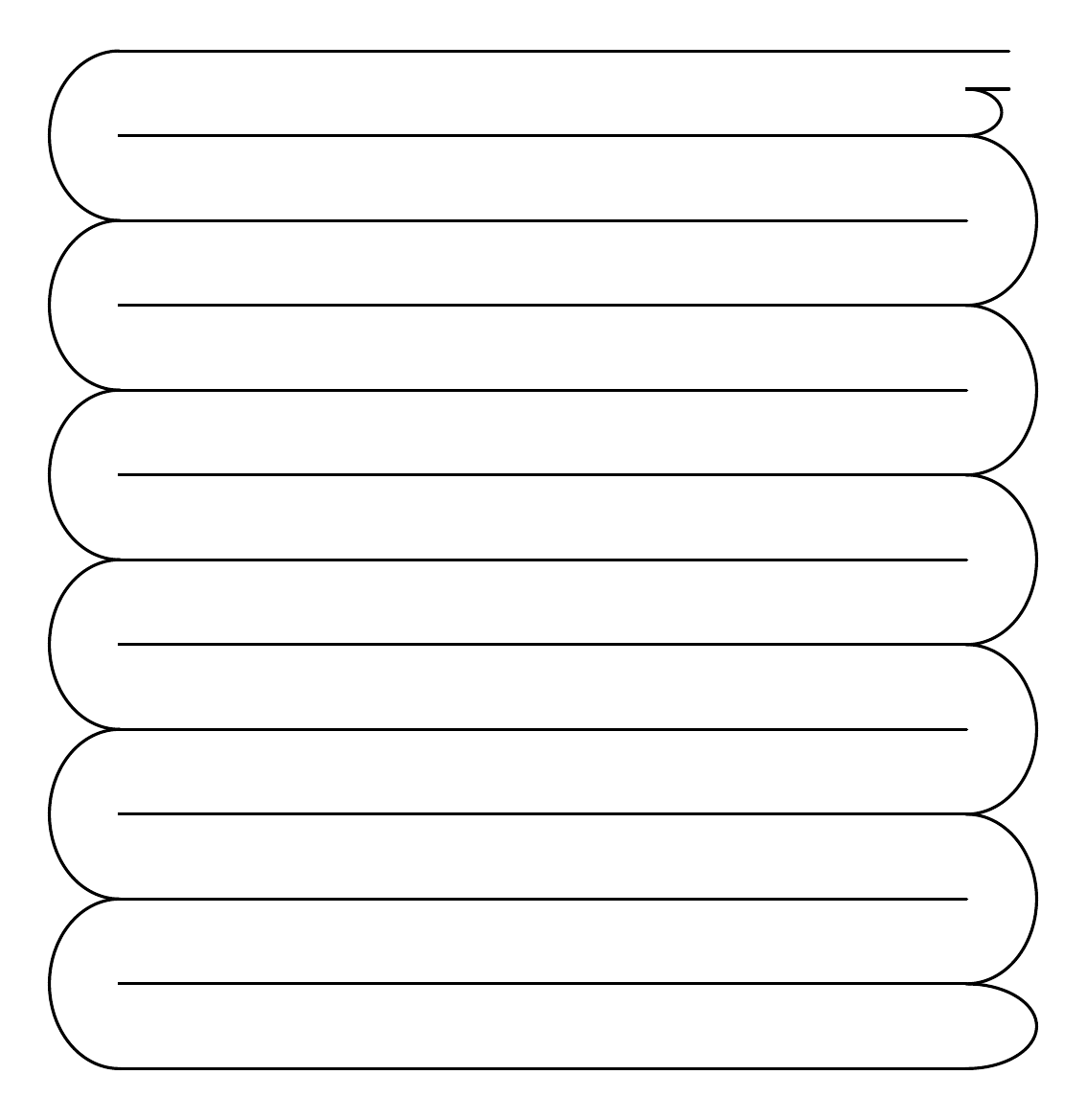}}
\caption{Array of 12 tubes with width \SI{15}{\milli\meter} and length \SI{150}{\milli\meter}.
}\label{app:svg}
\end{figure}

As it is parameterized, the script generates the pattern shown in Fig.~\ref{app:svg}. The input channel is designed for a luer lock input of size 3/32 inches -- note that the actual width of this channel depends on the soldering parameters.

\newpage

\paragraph*{Solving Kirchhoff rod equations.} 
We present here a Python script used to solve the coupled Kirchhoff rod system discussed in section C as a boundary-value problem. As parameterized, the problem is solved for $B_+/B_- =2$ and for dimensionless pressures $P^*$ varying between 0.1 and 1000.

\begin{lstlisting}[language=Python, caption={Script to solve coupled pressurized Kirchhoff rods}, label={lst:kirchhoff_code}]
    import numpy as np
    from scipy.integrate import solve_bvp
    
    # computation parameters
    n_node = 100
    n_pressure_step = 100
    
    # adimensionned parameters
    p_ini = 0.1 # small adimensionned pressure
    p_max = 1000
    rho = 0.5 # defined as B_-/B_+
    
    # Y = [theta1 (0), kappa1 (1), Rx1 (2), Ry1 (3), x1 (4), y1 (5), theta2 (6), kappa2 (7), Rx2 (8), Ry2 (9), x2 (10), y2 (11)]
    def f(x, y): # 6 Kirchhoff equations + 6 kinematic equations
        return np.vstack([y[1,:], - y[3,:], y[3,:]*y[1,:], +p - y[2,:]*y[1,:], np.cos(y[0,:]), np.sin(y[0,:]), y[7,:], - y[9,:]/rho, y[9,:]*y[7,:], -p - y[8,:]*y[7,:], np.cos(y[6,:]), np.sin(y[6,:]) ] ) # du/dx = p, dp/dx = u
    
    def bc(y0, y1): # y0 corresponds to s*_\pm=0, y1 to s*_pm=0.5
        return np.array([y0[0], y0[4], y0[5], y0[3], y0[6], y0[9], y1[0]-y1[6], y1[4]-y1[10], y1[5]-y1[11], y1[2]+y1[8], y1[3]+y1[9], y1[1]+rho*y1[7]])

    # initialize a guess for solution from the initial low pressure
    # once found, the solution can be calculated on a very large number of points.
    x = np.linspace(0, 0.5, n_node)
    yini = np.ones((12, n_node))
    yini[0,:]= - p_ini* (x**3)/6
    yini[6,:]= + p_ini* (x**3)/6
    yini[1,:]= - p_ini * (x**2)/2
    yini[7,:]= p_ini * (x**2)/2
    yini[2,:]=0*x
    yini[8,:]=0*x
    yini[3,:]= p_ini*x
    yini[9,:]= - p_ini*x
    yini[4,:] = x
    yini[10,:] = x
    yini[5,:] = x*0
    yini[11,:] = x*0
    
    # first iteration of the solver
    p = p_ini
    sol = solve_bvp(f, bc, x, yini,verbose=1)
    for idx, p in enumerate(np.linspace(p_ini, p_max, num=n_pressure_step)):
        # initial solutions are taken as the solution of the previous step 
        sol = solve_bvp(f, bc, sol.x, sol.y, verbose=0)
        # from sol, we can extract any variable of the problem
        phi_jonction = sol.y[0,-1] # for example here the junction angle

    # we can draw the cross-section
    plt.figure()
    plt.plot(sol.y[4,:], sol.y[5,:], '-k', linewidth=2) # bottom part (superior bending stiffnes)
    plt.plot(sol.y[10,:], sol.y[11,:], '-k', linewidth=2) # top part (inferior bending stiffness)
\end{lstlisting}

\clearpage

\clearpage
\bibliographystyle{unsrt}
\bibliography{biblio}

\end{document}